\documentclass[11pt]{article}
\usepackage{a4}
\usepackage{times}
\usepackage{amssymb}

\def\beq{\begin{equation}}
\def\bea{\begin{eqnarray}}
\def\eeq{\end{equation}}
\def\eea{\end{eqnarray}}

\def\Z{$\mathbb{Z}$}
\def\SM{Standard Model}

\begin{document}
\title{\bf  Stabilising the supersymmetric Standard Model \\ on the $\mathbb{Z}_6'$ orientifold}
\maketitle
\vglue 0.35cm
\begin{center}
\author{\bf David Bailin \footnote
{d.bailin@sussex.ac.uk} \&  Alex Love \\}
\vglue 0.2cm
	{\it  Department of Physics \& Astronomy \\ University of Sussex\\}
{\it Brighton BN1 9QH, U.K. \\}
\baselineskip=12pt
\end{center}
\vglue 2.5cm

\begin{abstract}Four stacks of intersecting supersymmetric fractional D6-branes on the \Z$_6'$ orientifold  have previously
been used to construct consistent models having the spectrum of the supersymmetric \SM, including a single pair of Higgs doublets, plus
three right-chiral neutrino singlets. However,  various moduli, K\"ahler moduli and complex-structure moduli, twisted
and untwisted, remain unfixed.
Further, some of the Yukawa couplings needed to generated quark and lepton masses are forbidden by a residual
global  symmetry of the model. In this paper we study the stabilisation of moduli using background fluxes,
 and show that the moduli
may be stabilised within the K\"ahler cone. In principle, missing Yukawa couplings may be restored, albeit with a coupling that is suppressed by non-perturbative effects,  by the use  Euclidean
D2-branes that are pointlike in spacetime, {\it i.e.} E2-instantons. However, for the models under investigation, we show that this is {\em not} possible.
\end{abstract}
\newpage
\section{Introduction}
The attraction of using intersecting D6-branes in a bottom-up approach to constructing the \SM \  is by now well known
\cite{Lust:2004ks}, and indeed  models having just the spectrum of the \SM \  have been constructed \cite{Ibanez:2001nd, Blumenhagen:2001te}. The four stacks of D6-branes wrap 3-cycles of an orientifold $T^6/\Omega$,
where the six extra spatial dimensions are assumed to be compactified on a 6-torus $T^6$ and
 $\Omega$ is the world-sheet parity operator; the use of an orientifold is essential to avoid the appearance of
additional vector-like matter. However, { non}-supersymmetric intersecting-brane models lead to flavour-changing
neutral-current (FCNC) processes induced by stringy instantons
 that can only be suppressed to levels consistent with current bounds by choosing a high
 string scale, of order $10^4$ TeV, which in turn leads to fine tuning problems \cite{Abel:2003yh}.
 It is therefore natural, and in any case
of interest in its own right, to construct  intersecting-brane models of the supersymmetric \SM. A supersymmetric theory
is not obliged to have a low string scale, so the instanton-induced FCNC processes may be reduced to rates well below the
experimental bounds by choosing a sufficiently high string scale without inducing fine-tuning problems.

To construct a supersymmetric theory \cite{Blumenhagen:2002gw, Honecker:2003vq, Honecker:2004kb, Gmeiner:2007zz, Gmeiner:2008xq}, instead of $T^6$, one starts  with an orbifold $T^6/P$,
where $P$ is a point group
which acts as an automorphism of the lattice defining $T^6$; this has the added advantage of fixing (some of) the
complex structure moduli. An orientifold is then constructed as before by quotienting the orbifold
 with the action of the world-sheet
parity operator $\Omega$. In previous papers \cite{Bailin:2006zf, Bailin:2007va, Bailin:2008xx} we have studied orientifolds with the  point group
$P=\mathbb{Z}_6'$, and derived models having the spectrum of the supersymmetric \SM \  plus three right-chiral neutrinos.
The 6-torus factorises into three 2-tori $T^6=T^2_1 \times T^2_2 \times T^2_3$, with $T^2_k \ (k=1,2,3)$ parametrised by
the complex coordinate $z^k$. Then the generator $\theta$ of the point group $P=\mathbb{Z}_6'$ acts on $z^k$ as
\beq
\theta z^k=e^{2 \pi i v^k}z^k
\eeq
where
\beq
(v^1,v^2,v^3)=\frac{1}{6}(1,2,-3)
\eeq
This requires that $T^2_1$ and $T^2_2$ are $SU(3)$ root lattices, so that $\theta$ is an automorphism, and
this in turn fixes the complex structure moduli $U_{1,2}$ for $T^2_{1,2}$ to be $U_1=U_2=e^{i\pi/3}\equiv \alpha$.
(Note that the $G_2$ and $SU(3)$ root lattices are the same, contrary to our previous assertions.)
Since $\theta$ acts on $T^2_3$ as a reflection, $\theta z^3=-z^3$, its lattice is arbitrary. The embedding
$\mathcal{R}$ of $\Omega$ acts   antilinearly on all $z^k$ and we may choose the phases so that
\beq
\mathcal{R} z^k=\bar{z}^k \ (k=1,2,3)
\eeq
This too must be an automorphism of the lattice, and this requires  the fundamental domain of
 each torus $T^2_k$ to be in one of two orientations, denoted {\bf A} and {\bf B}, relative to the ${\rm Re} \ z^k$ axis.
 In the {\bf A} orientation of $T^2_1$ the basis vector $e_1=R_1$ is real,
whereas in the {\bf B} orientation $e_1=R_1e^{-i\pi/6}$; for both orientations the second basis vector $e_2=\alpha e_1$.
Similarly for the basis vectors $e_3$ and $e_4$ of $T^2_2$. For $T^2_3$, the
basis vector $e_5=R_5$ is real in both orientations, but the real part of the complex structure $U_3 \equiv e_6/e_5$
satisfies ${\rm Re} \ U_3=0$ in the {\bf A} orientation, and ${\rm Re} \ U_3=1/2$ in the {\bf B} orientation. Thus
$e_6=iR_5{\rm Im} \ U_3$ in {\bf A}, and $e_6=R_5(1/2+i{\rm Im} \ U_3)$ in {\bf B}.

The models having the spectrum  of the supersymmetric \SM, with which we are concerned
in this paper, arise only in the {\bf AAA} and {\bf BAA} orientations. They include four stacks  of (supersymmetric)
 fractional  D6-branes, each wrapping the three large spatial dimensions and a 3-cycle of the general form
\beq
\kappa = \frac{1}{2}\left( \Pi ^{\rm bulk}_{\kappa}+\Pi ^{\rm ex}_{\kappa} \right)
\eeq
where
\beq
\Pi ^{\rm bulk}_{\kappa}=\sum _{p=1,3,4,6}A^{\kappa}_p\rho _p
\eeq
is an untwisted point-group invariant bulk 3-cycle, and
\beq
\Pi ^{\rm ex}_{\kappa}=\sum _{j=1,4,5,6}(\alpha _j\epsilon _j+\tilde{\alpha} _j\tilde{\epsilon} _j)
\eeq
is an exceptional cycle.
The four basis bulk 3-cycles $\rho _p \ (p=1,3,4,6)$  and their bulk coefficients $A^{\kappa}_p$
are defined in \cite{Bailin:2006zf}, the latter being expressed in terms of the wrapping numbers $(n^{\kappa}_k,m^{\kappa}_k)$ of
the basis 1-cycles $\pi _{2k-1},\pi _{2k}$ of $T^2_k  \ (k=1,2,3)$.
$\theta ^3$ acts as a $\mathbb{Z}_2$ reflection in $T^2_1$ and $T^2_3$ and therefore has sixteen
fixed points at
\beq
f_{i,j}=\frac{1}{2}(\sigma _1e_1+\sigma_2e_2) \otimes \frac{1}{2}(\sigma_5e_5+\sigma_6e_6) \label{fij}
\eeq
where $\sigma_{1,2,5,6}=0,1 \bmod 2$, and we use  Honecker's \cite{Honecker:2003vq, Honecker:2004np} notation in which  $i,j=1,4,5,6$ correspond to the pairs
$(\sigma_1,\sigma_2)$ or $(\sigma_5,\sigma_6)$
\beq
1 \sim  (0,0), \quad 4\sim (1,0), \quad 5 \sim (0,1), \quad 6 \sim (1,1) \label{1456}
\eeq
The eight exceptional  3-cycles $\epsilon _j, \ \tilde{\epsilon}_j \ (j=1,4,5,6)$ and their coefficients
$\alpha ^{\kappa}_j,\ \tilde{\alpha}^{\kappa}_j$
 are defined (in \cite{Bailin:2006zf}) in terms of collapsed 2-cycles
at $f_{i,j}$ times a 1-cycle in $T^2_2$, the coefficients being determined by
the wrapping numbers $(n^{\kappa}_2,m^{\kappa}_2)$ of the basis 1-cycles $\pi _3, \ \pi _4$ of $T^2_2$.
Supersymmetry requires that the bulk part of the fractional brane passes through the fixed points associated with the exceptional piece. 
If, for example, $(n^{\kappa}_3,m^{\kappa}_3)=(1,0) \bmod 2$, then, depending on the choice of Wilson lines,  only the exceptional cycles with $\alpha _{1,4}, \tilde{\alpha} _{1,4}$ or $\alpha _{5,6}, \tilde{\alpha} _{5,6}$ non-zero are allowed; similarly, for the $(0,1)\bmod 2$ case, only  $\alpha _{1,5}, \tilde{\alpha} _{1,5}$ or $\alpha _{4,6}, \tilde{\alpha} _{4,6}$  may be non-zero, and for the $(1,1)\bmod 2$ case, only  $\alpha _{1,6}, \tilde{\alpha} _{1,6}$ or $\alpha _{4,5}, \tilde{\alpha} _{4,5}$  may be non-zero. 
Orientifold invariance requires that we also include D6-branes wrapping the orientifold image
$\kappa' \equiv \mathcal{R}\kappa$ of each 3-cycle $\kappa$, and the action of $\mathcal{R}$ on the basis 3-cycles
$\rho_p,\epsilon _j, \tilde{\epsilon}_j$ is also given in \cite{Bailin:2006zf}.
The precise form of the 3-cycles associated with the four stacks is given in \cite{Bailin:2007va, Bailin:2008xx}
 and need not concern us for the present.  D6-branes carry Ramond-Ramond (RR) charge and are coupled electrically
to the 7-form RR gauge potential $C_7$. So too is the O6-plane, a topological defect associated with the orientifold action
which has $-4$ units of RR charge.

The massive version of the effective supergravity describing compactified type IIA string theory in the presence of background fluxes has action \cite{DeWolfe:2005uu, Louis:2002ny}
\bea
S_{IIA}&=&\frac{1}{2\kappa_{10}^2}\int d^{ 10}x \ \sqrt{-g}
\left( e^{-2\phi}[\mathbb{R}+4(\partial \phi)^2-\frac{1}{2}|H_3|^2]
-[|F_2|^2+|F_4|^2+m_0^2]\right) \nonumber \\
&-&\frac{1}{2\kappa_{10}^2}\int \left(B_2 \wedge dC_3\wedge dC_3+2B_2\wedge dC_3 \wedge F_4^{bg}+C_3 \wedge H_3^{bg}\wedge dC_3 \right. \nonumber \\
&&\left.-\frac{m_0}{3}B_2 \wedge B_2 \wedge B_2 \wedge dC_3+ \frac{m_0^2}{20}B_2 \wedge B_2 \wedge B_2 \wedge B_2 \wedge B_2 \right) \nonumber \\
&&-\mu _6 \sum _{\kappa}N_{\kappa}\int_{\mathcal{M}_4 \times\kappa}d^7\xi \ e^{-\phi}\sqrt{-g}
 + \sqrt{2} \mu _6\sum _{\kappa}N_{\kappa}\int_{\mathcal{M}_4 \times\kappa} C_7 \label{SIIA}
\eea
where  $2\kappa_{10}^2=(2\pi)^7 {\alpha'}^4$ is the 10-dimensional Newtonian gravitational constant and
$\mu _6=(2\pi)^{-6}{\alpha '}^{-7/2}$ is the unit of D6-brane RR charge.
The sum over $\kappa$ is understood to include all D6-brane stacks, their orientifold images $\kappa '$,
 and the ${\rm O6}$-brane $\pi _{\rm O6}$ with charge $-4 \mu _6$, and
$N_{\kappa}$ is the number of D6-branes in the stack wrapping the 3-cycle $\kappa$.
 The field strengths associated with the Kalb-Ramond field
$B_2$ and the RR fields $C_{1,3}$ are
 \bea
H_3&=&dB_2+H_3^{bg} \\
F_2&=&dC_1+m_0B_2  \label{F2}\\
F_4&=&dC_3+F_4^{bg}-C_1 \wedge H_3 -\frac{m_0}{2}B_2 \wedge B_2 \label{F4}
\eea
where $H_3^{bg}$ and $F_4^{bg}$ are background fluxes, and the mass $m_0$ is the background value of $F_0$.
The presence of the fluxes generally deforms the original metric. The direct product of the four-dimensional Minkowski space and the compactified (Calabi-Yau) space is replaced by a warped product \cite{Strominger:1986uh, de Wit:1986xg}
 which, as we shall see, introduces a potential for (some of) the moduli.
$dC_1$ is the Hodge dual of $F_8$, the field strength associated with the 7-form RR gauge field $C_7$.
One effect of the $m_0$ term is that a piece of the $F_2 \wedge ^*\!\!F_2$ term in (\ref{SIIA})
couples   $H_3^{bg}$  to  $C_7$
\beq
F_2 \wedge ^*\!\!F_2 \supset m_0H^{bg}_3 \wedge C_7
\eeq
so that this term also contributes to the $C_7$ tadpole equation. The requirement that there are no RR $C_7$ tadpoles is
therefore generalised  \cite{Camara:2005dc} to
\beq
\mu_6\left(\sum_{\kappa}N_{\kappa}(\kappa+ \kappa')-4\Pi_{\rm O6}\right)+\frac{1}{4\kappa _{10}^2}\Pi _{m_0H_3^{bg}}=0 \label{RR}
\eeq
where $\Pi _{m_0H_3^{bg}}$ is the 3-cycle of which $m_0H_3^{bg}$ is the Poincar\'e dual.
  In the models presented in  \cite{Bailin:2007va, Bailin:2008xx} tadpole cancellation requires that
$\Pi _{m_0H_3^{bg}}$, and hence $m_0H_3^{bg}$, is non-zero.

In general, we must also address the question of whether the total K-theory charge \cite{Witten:1998cd}  is zero. The presence of K-theory charge may be exhibited by the introduction of a ``probe'' $Sp(2) \simeq SU(2)$ brane $\pi _{\rm probe}$. For a consistent theory we require that there are an {\em even} number of chiral fermions in the fundamental representation of $Sp(2)$. Thus the additional constraint \cite{Gmeiner:2007we, Gmeiner:2007zz} is that
\beq
\sum _{\kappa}N_{\kappa}\kappa \cap \pi _{\rm probe}= 0 \bmod 2 \label{K02}
\eeq
where the sum is over all D6-branes, but {\em not} including their orientifold images, and $\pi_{\rm probe}$ is any 3-cycle 
that is its own orientifold image
\beq
\pi _{\rm probe}={\pi _{\rm probe}}'
\eeq
 although this may be too strong a constraint.
It follows that \cite{Gmeiner:2007we, Gmeiner:2007zz}
\beq
\pi _{\rm probe}= \frac{1}{2}\left( \Pi _{\rm probe}^{\rm bulk}+\Pi _{\rm probe}^{\rm ex} \right)
\eeq
where, on the {\bf AAA} lattice,
\bea
\Pi _{\rm probe}^{\rm bulk}=A_1\rho _1+A_4(\rho _4+2\rho_6) \\
\Pi _{\rm probe}^{\rm ex}=\sum _{j=1,4,5,6}\tilde{\alpha}_j(2\epsilon _j+\tilde{\epsilon}_j)
\eea
The two independent (supersymmetric) possibilities are
\bea
A_p&=&( 1,0,0,0) \quad \tilde{\alpha}_j=t_0(1,t_2,0,0) \quad {\rm or} \quad t_0(0,0,1,t_2) \label{A40}\\
{\rm or} \quad A_p&=&(0,0,1,2) \quad \tilde{\alpha}_j=t_0(1,0,t_2,0) \quad {\rm or} \quad t_0(0,1,0,t_2) \label{A402}
\eea
with $t_0,t_2=\pm 1$. In our models, in particular in the model deriving from the fourth entry in Table 1 of reference \cite{Bailin:2008xx}, the contributions to the left-hand side of (\ref{K02}) from the stacks $b$ and $c$ are necessarily {\em even}, the former because $N_b=2$, and the latter because it is zero. For the remaining stacks, we find that $a \cap \pi_{\rm probe}=-d \cap \pi_{\rm probe}$ for both cases (\ref{A40}) and (\ref{A402}) above. Thus the K-theory constraint (\ref{K02}) {\em is} satisfied. The same is true of the other models on the {\bf AAA} lattice,  as well as for the {\bf BAA} cases too.


All of the models that we have considered have the attractive feature that they have the spectrum of the supersymmetric \SM, including a single pair of Higgs doublets, plus
three right-chiral neutrino singlets. In the presence of these suitably chosen background fields $m_0$ and $H_3^{bg}$ the models {\em are}
consistent string theory vacua. Nevertheless, despite the attraction of having ``realistic'' spectra,
they are deficient. First, there are many unfixed moduli, K\"ahler moduli, complex structure moduli, axions and the dilaton,
all of which have unobserved massless quanta unless they are stabilised.  We shall see later that the non-zero background
flux $H_3^{bg}$ required by tadpole cancellation stabilises one linear combination of the (axion) moduli.
Tadpole cancellation generally ensures the absence of anomalous $U(1)$ gauge symmetries in the models;
 the associated gauge boson
acquires a string-scale mass via the generalised Green-Schwarz mechanism, and the $U(1)$ survives only as a global symmetry.
 However, some of the surviving global symmetries forbid the Yukawa couplings
required to generate mass terms for some of the quarks and leptons. This is the second deficiency of these models. 
Further, as noted previously in \cite{Bailin:2007va}, there is a surviving unwanted $U(1)_{B-L}$ gauge symmetry, associated with baryon number $B$ minus lepton number $L$. In addition, 
in all of the models that we constructed, the $U(1)$ stack associated with the fractional 3-cycle $c$ has the property that $c=c'$, where $c'$ is the orientifold image of $c$. This means that the $U(1)_c$ gauge symmetry is enhanced to $SP(2)=SU(2)$, so that the models  actually have as surviving gauge symmetry group $SU(3)_{\rm colour} \times SU(2)_L \times SU(2)_R \times U(1)_{B-L}$. 
The weak hypercharge is given by $Y=\frac{1}{2}(B-L)+T^3_R$, and the matter is in the following representations 
$({\bf n}_3,{\bf n}_L, {\bf n}_R)_{B-L}$ of $SU(3)_{\rm colour} \times SU(2)_L \times SU(2)_R \times U(1)_{B-L}$:
\bea
Q_L&=& ({\bf 3}, {\bf 2}, {\bf 1})_{\frac{1}{3}} \\
q^c_L&=&(\bar{\bf 3}, {\bf 1}, {\bf 2})_{-\frac{1}{3}} \\
L&=&({\bf 1}, {\bf 2}, {\bf 1})_{-1} \\
\ell^c_L, \nu^c_L&=&({\bf 1}, {\bf 1}, {\bf 2})_{1} \\
H_{u,d}&=&({\bf 1}, {\bf 2}, {\bf 2})_{0} 
\eea
In addition, the models we have constructed cannot yield gauge coupling constant unification.
A stack $\kappa$ gives rise to a gauge group factor with coupling constant $g_{\kappa}$ given  \cite{Klebanov:2003my,Blumenhagen:2003jy} by
\beq
\frac{1}{\alpha _{\kappa}} \equiv \frac{4\pi}{g_{\kappa}^2}=\frac{m_{\rm string}^3{\rm Vol}(\kappa)}{(2\pi)^3 g_{\rm string} K_{\kappa}} \label{alphkap}
\eeq
where ${\rm Vol}(\kappa)$ is the volume of the 3-cycle $\kappa$ and $K_{\kappa}=1$ for a $U(N_{\kappa})$ stack. The consistency of our treatment with the supergravity approximation requires that the  contribution of the bulk part of the fractional 3-cycle $\frac{1}{2}{\rm Vol}(\Pi _{\kappa} ^{\rm bulk})$  to ${\rm Vol}(\kappa)$ is large compared to the  contribution from the exceptional part $\frac{1}{2}{\rm Vol}(\Pi _{\kappa} ^{\rm ex})$, so we need only consider the former in evaluation $g_{\kappa}^2$. As derived in \cite{Bailin:2006zf}, for a supersymmetric stack, the quantity
\beq
Z^{\kappa} =  e_1e_3e_5 [A^{\kappa}_1-A^{\kappa}_3+U_3(A^{\kappa}_4-A^{\kappa}_6)+e^{i\pi/3}(A^{\kappa}_3+A^{\kappa}_6U_3)]>0 \label{Zkap}
\eeq
is real and positive. Here $A^{\kappa}_p \ (p=1,3,4,6)$ are the bulk wrapping numbers, $e_{2k-1} \ (k=1,2,3)$ are the basis vectors of $T^2_k$, and $U_3$ is the complex structure of $T^2_3$; the complex structure of $T^2_{1,2}$ is fixed by the \Z$_6'$ orbifold symmetry to be $U_{1,2}=e^{i\pi/3}$. Then
\beq
\frac{{\rm Vol}(\kappa)}{\sqrt{2{\rm Vol}(T^6/\mathbb{Z}_6')}}= \frac{Z^{\kappa}}{|e_1e_3e_5|\sqrt{|{\rm Im} \ U_3}|} \label{Volkap}
\eeq
The solutions for the {\bf AAA} lattice given in Table 1 of \cite{Bailin:2008xx}, in which $U_3=-i/\sqrt{3}$, all have
\beq
Z^a=2 |e_1e_3e_5| \quad {\rm and} \quad  Z^b=|e_1e_3e_5|
\eeq
Using equation (\ref{alphkap}) above, it follows that at the string scale $m_{\rm string}$ the coupling strengths for the $SU(3)_{\rm colour}$ and $SU(2)_L$ groups satisfy
\beq
\frac{\alpha _3}{\alpha _2}=\frac{1}{2}
\eeq
which is clearly inconsistent with the ``observed'' unification $\alpha _3=\alpha _2$ at the scale $m_X \simeq 2 \times 10^{16}$ GeV. We reach the same conclusion for the solutions on the {\bf BAA} lattice given in Table 6 of \cite{Bailin:2008xx}, in which $U_3=-i\sqrt{3}$. Thus, running from the string scale to the TeV scale with the three-generation supersymmetric Standard Model spectrum,
none of our solutions can reproduce the measured values of the non-abelian coupling strengths of the $SU(3)_{\rm colour}$ and $SU(2)_L$ gauge groups.
In fact the only supersymmetric models  obtained in \cite{Bailin:2006zf} yielding three chiral generations $3Q_L$ of quark doublets via $(a \cap b,a \cap b') = (2,1)$ or $(1,2)$, having no chiral matter in symmetric representations of the gauge groups, and not too much in antisymmetric representations, that also produce non-abelian coupling constant unification, 
 are the two solutions on the {\bf BAB} lattice given in Table 15 of that paper.  
We showed in \cite{Bailin:2008xx} that neither model can have just the required Standard Model spectrum, but it is of interest to see what can be achieved if we relax this constraint and allow additional vector-like
 matter but {\em not} extra chiral exotics. This requires at least two $U(1)$ stacks (both of which must be $d$-type in the terminology of that paper). The best we can do yields two additional vector-like Higgs doublets $2(H_u+H_d)$ and four additional vector-like charged lepton singlets $4(\ell^c_L+\bar{\ell}^c_L)$, and in any case the weak hypercharge $U(1)_Y$  gauge coupling strength $\alpha _Y\neq 3\alpha _3/5$ as required by the ``observed'' standard-model unification. We have not pursued this any further. The one-loop gauge threshold corrections to (\ref{alphkap}) have been computed by Gmeiner and Honecker \cite{Gmeiner:2009fb}. However, for the models under consideration, these are very small and the above conclusion is unaffected.
Another possibility that in principle might yield a realistic model is to start with an $SU(3)_{\rm colour}$ stack $a$ and an $SU(2)_L$ stack $b$ satisfying $(a \cap b, a \cap b')= (3,0)$ or $(0,3)$, and to require gauge coupling constant  unification $\alpha _3= \alpha _2$.  Following the work of Gmeiner and Honecker \cite{Gmeiner:2008xq}, we know at the outset that there are no such models that yield the standard-model spectrum {\em and} satisfy tadpole cancellation without the introduction of non-zero background flux $H_3^{bg}$.   However, since we have entertained the presence of such flux, it is of interest to know how far one can get with such models. We have searched for solutions satisfying both of these criteria, but have found none.

Finally, the presence of a non-zero flux $H_3^{bg}$ means that there may also arise a Freed-Witten anomaly \cite{Freed:1999vc}. In the presence of D6-branes the localised Bianchi identity associated with the stack $\kappa$ imposes the constraint \cite{Villadoro:2007tb}
\beq
H_3^{bg} \wedge [\kappa]=0
\eeq
where $[\kappa]$ is the 3-form that is the Poincar\'e dual of $\kappa$. Since $H_3^{bg}$ is odd under the orientifold action $\mathcal{R}$, only the $\mathcal{R}$-even part of $[\kappa]$, deriving from the $\mathcal{R}$-odd part of $\kappa$, can contribute to the anomaly. We have studied this in Appendix A. Our conclusion in all cases is that there is a non-zero anomaly deriving from the $SU(3)$ stack $a$ and also from one of the $U(1)$ stacks.

The deficiencies  detailed above mean that our models can only be considered as semi-realistic. Nevertheless, 
 it is of interest to see the extent to which the first two deficiencies can be remedied in models with a realistic spectrum. In this paper we study the fixing  of moduli using  background fluxes, the stability of these solutions and their consistency with the supergravity approximation in which they are derived. We also investigate the utility of non-perturbative effects,  so-called E2-instantons, to stabilise axion moduli and to repair the missing Yukawa couplings. 
\section{Moduli stabilisation} \label{susy}
In this and the following section we parallel the the treatment given by DeWolfe {\it et al.} \cite{DeWolfe:2005uu} of the
$\mathbb{Z}_3 \times \mathbb{Z}_3$ orientifold.
It has been shown by Grimm and Louis \cite{Grimm:2004ua} that the effective four-dimensional theory deriving from type IIA supergravity
compactified on a Calabi-Yau 3-fold is an $\mathcal{N}=2$ supergravity theory.
The moduli space is the product of two factors, one containing the vector multiplets (which include the K\"ahler moduli),
and the other the hypermutiplets  (which include the complex structure moduli and dilaton). The metric on each space is
derived from a K\"ahler potential, $K^{K}$ and $K^{cs}$ respectively. The orientifold projection $\mathcal{R}$ to an
 $\mathcal{N}=1$ supergravity reduces the size of each moduli space.

Consider first the K\"ahler moduli. The complexified K\"ahler form
\beq
J_c=B_2+iJ
\eeq
 is {\em odd} under the action of
$\mathcal{R}$ and can therefore be expanded in terms of the   $\mathcal{R}$-odd $(1,1)$-forms.
In our case, on the $\mathbb{Z}_6'$ orbifold,  we have three  untwisted, invariant $(1,1)$-forms $w_k \ (k=1,2,3)$ defined by
\beq
w_k \equiv dz^k \wedge d\bar{z}^k \quad ({\rm no \ summation}) \label{wk}
\eeq
There are also eight $\theta ^3$-twisted sector invariant harmonic $(1,1)$-forms $e_{(1,j)}, \ \hat{w}_j, \ (j=1,4,5,6)$,
 defined as follows. Associated with each of the 16 fixed points $f_{i,j}$, defined in (\ref{fij}),  is a
localised $(1,1)$-form
\beq
e_{(i,j)} \equiv \omega_{k,\bar{\ell}}dz^k \wedge d\bar{z}^{\ell} \quad (k,\ell=1,3) \label{eij}
\eeq
After blowing up  the fixed point using the Eguchi-Hanson $EH_2$ metric \cite{Lutken}, $\omega_{k,\bar{\ell}}$ has the form
\beq
\omega _{k \bar{\ell}}=a(u)\delta _{k\bar{\ell}}+b(u)(z_k-Z_k)(\bar{z}_{\ell}-\bar{Z}_{\ell})
\eeq
when the fixed point $f_{i, j}$ is at $(z^1,z^3)=(Z^1,Z^3) \in T^2_1 \times T^2_3$. The functions $a(u)$ and $b(u)$ are
given by
\bea
a(u)&=&u^{-1}(\lambda^4+u^2)^{-1/2}\lambda ^4 \\
b(u)&=&a'(u)
\eea
with $\lambda $ the blow-up parameter and
\beq
u \equiv |z^1-Z^1|^2+ |z^3-Z^3|^2 \label{u}
\eeq
Under the action of the point group generator $\theta$ these $(1,1)$ forms transform as
\bea
&&e_{(1,j)} \rightarrow e_{(1,j)} \label{e1j} \\
&&e_{(4,j)} \rightarrow e_{(6,j)} \rightarrow e_{(5,j)} \rightarrow e_{(4,j)} \label{e456j}
\eea
Thus the eight invariant $\theta ^3$-twisted $(1,1)$ forms are $e_{(1,j)} $ and
\beq
\hat{w}_j \equiv  e_{(4,j)}+e_{(5,j)}+e_{(6,j)} \ (j=1,4,5,6)
\eeq
We denote the blow-up parameter associated with $e_{(1,j)} $ by $\lambda _j$. Point-group invariance (\ref{e456j}) requires that $e_{(4,j)}$, $e_{(5,j)}$ and $e_{(6,j)}$ all have the same blow-up parameter, which we denote by $\hat{\lambda}_j$. 
All of the invariant $\theta ^3$-twisted $(1,1)$ forms given above are odd under the action of $\mathcal{R}$, so in general we may expand the complexified K\"ahler form as
\beq
J_c=\sum_{k=1,2,3}t_kiw_k+\sum _{j=1,4,5,6}(T_jie_{(1,j)}+\hat{T}_ji\hat{w}_j) \label{Jc}
\eeq
where
\bea
t_k=b_k+iv_k  \label{tk} \\
T_j=B_j+iV_j \label{Tj}\\
\hat{T}_j=\hat{B}_j+i\hat{V}_j \label{hatTj}
\eea
$b_k,B_j,\hat{B}_j$ are associated with the Kalb-Ramond field $B_2$, and
the K\"ahler moduli $v_k,V_j,\hat{V}_j$ with the K\"ahler form $J$.
The K\"ahler potential $K^K$ for the K\"ahler moduli is given by
\bea
K^K&=&-\log \int \left( \frac{4}{3}\int_{T^6/\mathbb{Z}_6'} J \wedge J \wedge J \right) \\
&=&-\log \left(\frac{32}{3} {\rm Vol}_6v_1v_2v_3-16\pi^2{\rm Vol}_2\sum _{j}v_2(\lambda _j^4V_j^2+3\hat{\lambda}_j^4\hat{V}_j^2) \right)
\eea
where ${\rm Vol}_{6,2}$ are the coordinate volumes of $T^6$ and $T^2_2$ respectively.
Thus
\beq
{\rm Vol}_6 =\prod_{k=1,2,3}{\rm Vol}_k \label{Vol6}
\eeq
where
\beq
{\rm Vol}_k  = R_{2k-1}^2{\rm Im} U_k \label{Volk}
\eeq
As previously noted, the $SU(3)$ lattice used for $T^2_{1,2}$ has $U_1=\alpha=U_2$, so that
${\rm Im }U_1=\sqrt{3}/2={\rm Im }U_2$. For the models found in \cite{Bailin:2007va, Bailin:2008xx}, ${\rm Im }U_3=-1/\sqrt{3}$ on the
{\bf AAA} lattice and $-\sqrt{3}$ on the {\bf BAA} lattice. It is convenient to absorb the coordinate volumes into the
moduli, so  we make the redefinitions
\bea
t_k{\rm Vol}_k \rightarrow t_k \label{tknew}\\
T_j \pi \lambda _j^2 \rightarrow T_j \label{Tjnew}\\
\hat{T}_j \pi \hat{\lambda} _j^2 \rightarrow \hat{T}_j \label{hatTjnew}
\eea
and then
\beq
K^K=-\log \left(\frac{32}{3} v_1v_2v_3-16\sum _{j}v_2(V_j^2+3\hat{V}_j^2)\right)  \label{KK}
\eeq
Note that, unlike in the $\mathbb{Z}_3 \times \mathbb{Z}_3$ case discussed in  \cite{DeWolfe:2005uu}, the twisted moduli $V_j$ and $\hat{V}_j$ are inextricably coupled to the untwisted modulus $v_2$.

The complex  structure moduli are obtained by expanding the holomorphic $(3,0)$-form $\Omega$ in terms of the basis
3-forms. There are four $\mathbb{Z}_6'$-invariant untwisted 3-forms, defined as in \cite{Bailin:2008xx} by
\bea
&&\sigma _0 \equiv dz^1 \wedge dz^2 \wedge dz^3 \label{sig0}\\
&&\sigma _1 \equiv dz^1 \wedge dz^2 \wedge d\bar{z}^3 \\
&&\sigma _2 \equiv d\bar{z}^1 \wedge d\bar{z}^2 \wedge dz^3 =\overline{\sigma}_1\\
&&\sigma _3 \equiv d\bar{z}^1 \wedge d\bar{z}^2 \wedge d\bar{z}^3 =\overline{\sigma}_0 \label{sig3}
\eea
Hence
\bea
\mathcal{R}(\sigma _0 \pm \sigma_3)=\pm (\sigma _0 \pm \sigma_3)\\
\mathcal{R}(\sigma _1 \pm \sigma_2)=\pm (\sigma _1 \pm \sigma_2)
\eea
The invariant $\theta ^3$-twisted 3-forms $\omega _j, \tilde{\omega}_j \ (j=1,4,5,6)$ are also as defined in \cite{Bailin:2008xx} as
\bea
\omega _j & \equiv & [\alpha(e_{(4,j)}-e_{(5,j)})+(e_{(5,j)}-e_{(6,j)})]\wedge dz_2  \label{omj}\\
\tilde{\omega}_j&\equiv & [(e_{(4,j)}-e_{(5,j)})+\alpha(e_{(5,j)}-e_{(6,j)})] \wedge d\bar{z}_2 \label{tilomj}
\eea
 Then
\bea
\mathcal{R}(\omega _j \mp \alpha \tilde{\omega}_j)= \pm (\omega _j \mp \alpha \tilde{\omega}_j) \ {\rm on} \  {\bf AAA} \\
\mathcal{R}(\tilde{\omega} _j \mp \alpha{\omega}_j)= \pm (\tilde{\omega} _j \mp \alpha{\omega}_j) \ {\rm on} \ {\bf BAA}
\eea
As above, it is convenient to factorise out coordinate volumes, so that the K\"ahler potential $K^{\rm cs}$
for the complex structure moduli is independent of them. Then on the {\bf AAA} lattice we may expand the holomorphic 3-form as
\bea
 \Omega&=& \frac{1}{\sqrt{{\rm Vol}_6}}[Z_0(\sigma _0 +\sigma _3)-g_0(\sigma _0 -\sigma _3)+
 Z_1(\sigma _1 +\sigma _2)-g_1(\sigma _1 -\sigma _2)] \nonumber \\
&&+\sum_j\frac{1}{\pi \hat{\lambda} _j^2 \sqrt{{\rm Vol}_2}}[Y_j \alpha ^2(\omega _j-\alpha \tilde{\omega}_j)
-f_j\alpha ^2(\omega _j+\alpha \tilde{\omega}_j)] \label{Oma}
 \eea
 On the {\bf BAA } lattice  $\omega _j$ and $\tilde{\omega}_j$ are interchanged.
In both cases $Z_{0,1}$ and  $Y_j$ are associated with the $\mathcal{R}$-even forms, and
 $g_{0,1},f_j$ with the $\mathcal{R}$-odd ones.
 It is easy to show that the complex conjugates
of the twisted 3-forms are given by
 \bea
 \bar{\omega}_j=\alpha ^2\tilde{\omega}_j \\
  \bar{\tilde{\omega}}_j=\alpha ^2{\omega}_j
 \eea
 The orientifold constraint requires  that
 \beq
\mathcal{R}\Omega= \bar{\Omega}
\eeq
which gives
\beq
Z_{0,1}, \ g_{0,1}, \ Y_j, \ f_j \quad {\rm are \ real} \label{Zgreal}
\eeq
The required K\"ahler potential is
\bea
K^{\rm cs}&=&-\log \left( i\int_{T^6/\mathbb{Z}_6'} \Omega \wedge \bar{\Omega} \right) \\
&=&-\log \left(  - \frac{16}{3}(Z_0g_0-Z_1g_1)+48 \sum_jY_jf_j        \right) \label{Kcs}
\eea

The $\mathcal{R}$ projection projects out half of the moduli of the $\mathcal{N}=2$ theory, including  half of the
universal hypermultiplet; the dilaton and one axion survive. The surviving moduli are all contained in the
complexified 3-form
\beq
\Omega _c \equiv C_3 +2i {\rm Re}(C\Omega)
\eeq
where $C_3$ is the RR 3-form gauge potential, and $C$ is the ``compensator'' that incorporates the dilaton dependence
\beq
C \equiv e^{-D+K^{cs}/2} \label{C}
\eeq
with the four-dimensional dilaton $D$ defined by
\beq
e^D \equiv \sqrt{8}e^{\phi+K^K/2}
\eeq
Since $C_3$ is even under the action of $\mathcal{R}$ we may expand it as
\beq
C_3=\frac{1}{   \sqrt{{\rm Vol}_6}      }[x_0(\sigma_0+\sigma_3)+x_1(\sigma_1+\sigma_2)]+
\sum_j\frac{1}{    \pi \hat{\lambda} _j^2 \sqrt{{\rm Vol}_2 }      }X_j \alpha ^2(\omega _j-\alpha \tilde{\omega}_j) \label{C3}
\eeq
on the  {\bf AAA} lattice; as before, in the {\bf BAA} case we interchange $\omega _j$ and $\tilde{\omega}_j$.
Expanding $\Omega _c$ as in (\ref{Oma}), on the  {\bf AAA} lattice
\bea
\Omega _c&=&\frac{1}{   \sqrt{{\rm Vol}_6}    }[N_0(\sigma _0 +\sigma _3)-T_0(\sigma _0 -\sigma _3)+ N_1(\sigma _1 +\sigma _2)
-T_1(\sigma _1 -\sigma _2)]  \nonumber \\
&&+\sum_j\frac{1}{   \pi \lambda _j^2 \sqrt{{\rm Vol}_2 }   }     [M_j \alpha ^2(\omega _j-\alpha \tilde{\omega}_j)-
S_j\alpha ^2(\omega _j+\alpha \tilde{\omega}_j)]  \label{Omca}
 \eea
with the usual interchange for the {\bf BAA} case.
Then the surviving moduli are the expansion of $\Omega _c$ in $H^3_{+}$,
{\it i.e.} the $\mathcal{R}$-even states with moduli
\bea
N_k&=&x_k+2iCZ_k \quad (k=0,1)  \label{N0} \\
M_j&=&{{X}_j}+2iCY_j \quad (j=1,4,5,6) \label{Mj}
\eea
in both cases.

The potential $V$ arising after dimensionally reducing the massive type IIA supergravity is
\beq
V=e^K\left( \sum_{i,j=\{t_k,T_j,\hat{T}_j,N_k,M_j\}}K^{ij}  D_iW\overline{D_jW}  -3|W|^2
  \right)+m_0e^{K^Q}{\rm Im} \ W^Q
\eeq
where the K\"ahler potential $K=K^K+K^Q$ with
\beq
K^Q=-2\log \left( 2\int {\rm Re} (C\Omega) \wedge ^*\!\!{\rm Re}  (C\Omega) \right)
\eeq
It follows from (\ref{Oma}) that on the {\bf AAA} lattice
\beq
{\rm Re}(C\Omega)= \frac{1}{\sqrt{{\rm Vol}_6}}[CZ_0(\sigma _0 +\sigma _3)+C Z_1(\sigma _1 +\sigma _2)]+
\sum_j \frac{1}{\pi \hat{\lambda} _j^2 \sqrt{{\rm Vol}_2}}\alpha ^2(\omega _j-\alpha \tilde{\omega}_j)
 \eeq
Also, since $\Omega$ is the holomorphic $(3,0)$-form, $^*\Omega =-i\Omega$ and
\beq
^*{\rm Re}(C\Omega)={\rm Re} ^*\!(C\Omega)=\frac{1}{\sqrt{{\rm Vol}_6}}[iCg_0(\sigma _0 -\sigma _3)+
iC g_1(\sigma _1 +\sigma _2)]+i\sum_j\frac{1}{\pi \hat{\lambda} _j^2 \sqrt{{\rm Vol}_2}} g_j \alpha ^2(\omega _j+\alpha \tilde{\omega}_j)
 \eeq
 so that
 \beq
e^{-K^Q/2}=C^2e^{-K^{cs}} = e^{-2D} \label{KcsQD}
\eeq
 where $K^{cs}$ is given in (\ref{Kcs}), and the last equality follows
from the definition (\ref{C}). The same result follows on the {\bf BAA} lattice. Like the K\"ahler moduli $t_k$,  the complex structure
  moduli $N_{0,1}, M_j$ enter the K\"ahler potential only via their imaginary parts.
  The superpotential \cite{Derendinger:2004jn, Villadoro:2005cu, Derendinger:2005ph} is $W=W^Q+W^K$ where
\bea
W^Q(N_k,M_j)&\equiv& \int\Omega _c \wedge H_3^{bg} \label{WQ} \\
W^K(t_k,T_j,\hat{T}_j)&\equiv& e_0+\int J_c \wedge  F_4^{bg}-
\frac{1}{2}\int J_c\wedge J_c \wedge F_2^{bg} - \frac{m_0}{6}\int J_c \wedge J_c \wedge J_c \label{WK}
\eea
and
\beq
e_0 \equiv\int F_6^{bg}   \label{e0F6}
\eeq
We note that $W^Q$ depends only on the NS-NS flux $H_3^{bg}$ and $W^K$ only on the RR fluxes $F^{bg}_{n} \ (n=0,2,4,6)$.
$H_3$ is odd under the action of $\mathcal{R}$, so that, analogously to (\ref{C3}), we may expand its background value as
\beq
H_3^{bg}=\frac{i}{   \sqrt{{\rm Vol}_6}      }[p_0(\sigma_0-\sigma_3)+p_1(\sigma_1-\sigma_2)]+
\sum_j\frac{i}{    \pi\hat{ \lambda} _j^2 \sqrt{{\rm Vol}_2 }     }P_j \alpha ^2(\omega _j+\alpha \tilde{\omega}_j) \label{H3bg}
\eeq
on the {\bf AAA} lattice, with $\omega \leftrightarrow \tilde{\omega}$ on {\bf BAA}. As shown in \cite{Bailin:2008xx},
flux quantisation requires that the coefficients are quantised. On the {\bf AAA} lattice
\bea
(p_0,p_1)&=&-\frac{  \pi ^2 \alpha ' \sqrt{  {\rm Vol}_6}   }{   3 \sqrt{3}R_1R_3R_5}(n_3+3n_6,n_3-3n_6) \ {\rm with} \  n_3,n_6 \in \mathbb{Z} \label{p0p1a}\\
P_j&=&-\frac{   2\pi ^2 \alpha '\sqrt{{\rm Vol}_2 } }{3R_3         }\hat{n}_j \    {\rm with} \ \hat{n}_j \in \mathbb{Z} \label{Pja}
\eea
where $n_{3,6}$ and $\hat{n}_j$ respectively  are associated with the flux of $H_3^{bg}$  through the 3-cycles  $\rho_{3,6}$ and $\epsilon _j$; note that $p_{0,1},P_j$ are {\em independent} of the coordinate scales $R_{1,3,5}$.
For the solution discussed in \S 5.1 of reference \cite{Bailin:2008xx}, relating to the fourth solution in Table 1 of that paper, 
the exceptional part of the tadpole cancellation  condition (\ref{RR}) requires that $|n_0\hat{n}_j|=12$ for $j=4,6$. Thus $|n_0|=1,2,3,4,6,12$. For $j=1$, we get $|n_0\hat{n}_1|=12|1-t^c_1|=0,24$, which is always consistent with these values of $n_0$.  Cancellation of the untwisted part proportional to $\rho_4+2\rho_6$ requires that the corresponding values of $n_3$ satisfy $|n_3|=1296,648,432,324,216,108$, and of $n_6$ satisfy $|n_6|=(1+t^c_1)(144,72,48,36,24,12)$. 
($t^c_1=\pm 1$ is one of the Wilson lines associated with the stack $c$.)

Alternatively, on the {\bf BAA} lattice
\bea
(p_0,p_1)&=&\frac{  \pi ^2 \alpha ' \sqrt{{\rm Vol}_6}   }{   9 R_1R_3R_5}(n_4+3n_1,-n_4+3n_1) \ {\rm with} \  n_1,n_4 \in \mathbb{Z} \\
P_j&=&\frac{   2\pi ^2 \alpha 'R_3  }{    \sqrt{  3{\rm Vol}_2   }     }\tilde{n}_j \    {\rm with} \ \tilde{n}_j \in \mathbb{Z} \label{Pjb}
\eea
where $n_{1,4}$ and $\tilde{n}_j$ respectively  are associated with the flux of $H_3^{bg}$  through the 3-cycles  $\rho_{1,4}$ and $\tilde{\epsilon }_j$.
In this case tadpole  cancellation of the exceptional parts requires that
 $| n_0\tilde{n}_j|=12$, so that $|n_0|=1,2,3,4,6,12$. Then, $n_4=0$ and the corresponding values of $n_1$ satisfy $|n_1|=432,216,144,108,72,36$. 

The form (\ref{H3bg}) for $H_3^{bg}$ gives
\beq
W^Q(N_k,M_j)=-\frac{8}{3}(N_0p_0-N_1p_1)+24\sum_j M_jP_j \label{WQ2}
\eeq
The background fluxes $F_2^{bg}$ and $F_4^{bg}$ that appear in $W^K$ have similar expansions. Since $F_2$ is odd under
the action of $\mathcal{R}$ and $F_4$ even
\bea
F_2^{bg}&=& \sum_{k=1,2,3}\frac{1}{{\rm Vol}_k}f_k iw_k+\sum _{j=1,4,5,6}\left(\frac{F_j}{\pi \lambda _j^2}ie_{(1,j)}+\frac{\hat{F}_j}{\pi\hat{\lambda}_j^2}i\hat{w}_j\right) \\
F_4^{bg} &=& \frac{1}{{\rm Vol}_6} \sum_{k=1,2,3}{\rm Vol}_ke_k\tilde{w}_k+
\sum _{j=1,4,5,6}\left(\frac{E_j}{{\rm Vol}_2 \pi \lambda _j^2}w_2 \wedge e_{(1,j)}+\frac{\hat{E}_j}{{\rm Vol}_2 \pi \hat{\lambda} _j^2}w_2 \wedge \hat{w}_j\right)  \nonumber \\
&+& \sum _{j=1,4,5,6}\left(\frac{G_j}{ \pi^2 \lambda _j^4} e_{(1,j)} \wedge e_{(1,j)}+\frac{\hat{G}_j}{\pi^2 \hat{\lambda} _j^4} \hat{w}_j \wedge \hat{w}_j\right) \label{F4bg}
\eea
where
\beq
\tilde{w}_k= dz^i \wedge d\bar{z}^i \wedge dz^j \wedge d\bar{z}^j \quad {\rm where} \quad (i,j,k)={\rm cyclic} \ (1,2,3)
\eeq
The constant term $e_0$ in $W^K$, defined in (\ref{e0F6}), arises from the Hodge dual $F_6^{bg}$ of $F_4$ polarised in the {\em non}-compact directions.
All of these fluxes, including $F_6^{bg}$,   are quantised, the general constraint being that for any closed $(p+2)$-cycle $\Sigma _{p+2}$
\beq
\mu _p\int_{\Sigma_{p+2}}F_{p+2}=2\pi n \quad {\rm with} \quad n \in \mathbb{Z}
\eeq
 with $\mu_p=(2\pi )^{-p} {\alpha '}^{-(p+1)/2}$ the electric charge of a D$p$-brane.
For the present, we set $F_2^{bg}=0$, and then
\bea
W^K(t_k,T_j,\hat{T}_j)&=&e_0-\frac{4}{3} \sum _{k=1}^3 t_ke_k+4\sum _j (T_{j}E_{j}+3\hat{T}_j\hat{E}_j) 
+4t_2\sum _j(G_j +3 \hat{G}_j) \nonumber\\
&&-m_0\left(\frac{4}{3}t_1t_2t_3-2\sum _{j}t_2(T_j^2+3\hat{T}_j^2)\right) \label{WK2}
\eea

The advantage of this formalism is that we may immediately identify supersymmetric vacua by their vanishing $F$-terms:
\beq
F_i =D_iW \equiv \partial _iW + W \partial _iK=0 \label{Fi}
\eeq
for every chiral superfield $i$.  For the complex-structure moduli, taking $i=N_k,M_j$, we get
 \bea
&&p_k+2ie^{2D}W(Cg_k)=0 \quad (k=0,1)  \label{pgk} \\
&&P_j+2ie^{2D}W(Cf_j)=0 \quad (j=1,4,5,6) \label{PFj}
\eea
As in \cite{DeWolfe:2005uu}, the imaginary parts  of these equations are degenerate.
Using (\ref{WQ}) and (\ref{N0}) ... (\ref{Mj}), they give the single constraint
\beq
{\rm Re} \ W=0 \label{ReW}
\eeq
which fixes only {\em one} linear  combination of the axions $x_0,x_1$ and $X_j$
\beq
\frac{8}{3}(x_0p_0-x_1p_1)-24\sum _jX_j{P}_j ={\rm Re} W^K \label{axion}
\eeq
This degeneracy derives from the fact that the coeffcients $p_k,P_j$ that determine $H_3^{bg}$ are {\em real}, and therefore have insufficient degrees of freedom to stabilise both the complex structure moduli and their axionic partners. As we shall discuss later, in \S \ref{se2}, E2-instantons can lift  the remaining degeneracy. 
The real parts give
\beq
e^{-K^{\rm cs}/2}\frac{p_k}{g_k}=e^{-K^{\rm cs}/2}\frac{P_j}{f_j}=Q_0 \label{real}
\eeq
where
\beq
Q_0 \equiv {\rm Im}We^D \label{Q0}
\eeq
Then (\ref{real}) determines the moduli $g_k, f_j$ up to an overall scale fixed by $Q_0$. Finally, using (\ref{KcsQD}),
(\ref{Q0}) gives
\beq
e^{-\phi}=\sqrt{8}e^{K^K/2}\frac{{\rm Im} W}{Q_0} \label{ephi}
\eeq
which fixes the dilaton once the other moduli are all fixed \cite{DeWolfe:2005uu}.
It follows from (\ref{pgk}), (\ref{PFj}), (\ref{Kcs}), (\ref{KcsQD}) and (\ref{WQ2}) that
\beq
{\rm Im} W^Q+2iW=0
\eeq
Thus, using  (\ref{ReW}), when the complex structure moduli satisfy their field equations
\beq
2{\rm Im} W^K+{\rm Im} W^Q=0
\eeq
and the vacuum value of the superpotential is determined entirely by the K\"ahler moduli
\beq
W=-i {\rm Im} W^K(t_k,T_j,\hat{T}_j) \label{ImWK}
\eeq

Vanishing F-terms for the K\"ahler moduli in (\ref{Fi}) give
\bea
e_k+m_0\frac{t_1t_2t_3}{t_k}-4iWe^{K^K}\frac{v_1v_2v_3}{v_k}&=&\frac{3}{2}\delta_{k2} \sum_j \left[m_0(T_j^2+3\hat{T}_j^2) \right. \nonumber \\
&+& \left. 2(G_j+3\hat{G}_j) -4iWe^{K^K}(V_j^2+3\hat{V}_j^2)\right] \\
E_j+m_0t_2T_j&=&4iWe^{K^K}v_2V_j \\
\hat{E}_j+m_0t_2\hat{T}_j&=&4iWe^{K^K}v_2\hat{V}_j
\eea
Using (\ref{ReW}), the imaginary parts of these equations require that
\beq
{\rm Im}(\partial _i W^K)=0 \quad {\rm for} \quad i=t_k,T_j,\hat{T}_j
\eeq
The simplest solution of these is
\beq
b_k=0=B_j=\hat{B}_j \label{B0}
\eeq
and then the above equations reduce to
\bea
e_k&=&\frac{3}{2}\delta_{k2} \sum_j [2(G_j+3\hat{G}_j)-X(V_j^2+3\hat{V}_j^2)]+X\frac{v_1v_2v_3}{v_k}\\
E_j&=&Xv_2V_j\\
\hat{E}_j&=&Xv_2\hat{V}_j
\eea
where
\beq
X \equiv m_0+4iWe^{K^K}=m_0+4{\rm Im}W^Ke^{K^K} \label{X}
\eeq
using (\ref{ImWK}).
They couple the untwisted volume modulus $v_2$ to the twisted volume moduli $V_j,\hat{V}_j$. Solving for all moduli in terms of $v_2$ and $X$ gives
\bea
v_1&=&\frac{e_3}{Xv_2}  \label{v1}\\
v_3&=&\frac{e_1}{Xv_2}  \label{v2}\\
V_j&=&\frac{E_j}{Xv_2}\\
\hat{V}_j&=&\frac{\hat{E}_j}{Xv_2} \label{hatVj}
\eea
Substituting these into the $v_2$ equation gives
\beq
\tilde{e}_2Xv_2^2=e_1e_3-\frac{3}{2}\sum_j(E_j^2+3\hat{E}_j^2) \equiv F(e_k,E_j,\hat{E}_j)
\eeq
where
\beq
\tilde{e}_2 \equiv e_2-3\sum_j(G_j+3\hat{G}_j)
\eeq
Then (\ref{KK}) gives
\beq
e^{K^K}=\frac{3X}{32\tilde{e}_2v_2}
\eeq
and the definition (\ref{X})  yields
\beq
\frac{3{\rm Im}W^K}{8\tilde{e}_2v_2}=1-\frac{m_0}{X}
\eeq
Substituting (\ref{v1}) ...(\ref{hatVj}) and (\ref{B0}) into (\ref{WK2}), it then follows that when the K\"ahler moduli satisfy their field equations,
\beq
X=\frac{3}{5}m_0
\eeq
so that
\beq
|v_2|=\sqrt{\frac{5F(e_k,E_j,\hat{E}_j)}{3\tilde{e}_2m_0}}
\eeq
Thus the background fluxes $e_k,E_j,\hat{E}_j, G_j, \hat{G}_j$ and $m_0$ fix $v_2$ and $X$, and hence, via equations  (\ref{v1}) ...(\ref{hatVj}), the remaining K\"ahler moduli.

The effective supergravity theory is a justifiable approximation \cite{DeWolfe:2005uu} so long as the volumes $v_k, V_j, \hat{V}_j$ are large enough that the  ${\rm O}(\alpha')$ corrections are negligible and the string coupling $g_s$ is small enough to neglect  corrections. Further, to remain within the K\"ahler cone we require that the untwisted volumes are large compared with the blow-up volumes, {\it i.e.} $v_k \gg V_j,\hat{V}_j\gg 1$. Since (the non-zero value of)  $m_0$ is fixed by the RR tadpole cancellation condition (\ref{RR}), and we have set $F_2^{bg}=0$, the question then is whether there are choices of the background 4-form flux $F_4^{bg}$  for which these constraints are obeyed. 

It follows from equations (\ref{v1}) ... (\ref{hatVj})  that  $v_{1,3}/V_j=e_{1,3}/E_j$, so that the K\"ahler cone constraints require that $e_1,e_3 \gg E_j$, and similarly for $\hat{E}_j$. Hence $F\sim e_1e_3$.
Then the constraints $v_k \gg 1$ require that $e_1e_3 \gg \tilde{e}_2m_0$, $e_1\tilde{e}_2 \gg e_3m_0$ and $e_3\tilde{e}_2 \gg e_1m_0$,  and these imply that $e_1,\tilde{e}_2, e_3 \gg m_0$. For the blow-up volumes, similarly, the constraints $v_k \gg V_j,\hat{V}_j \gg 1$ require that $e_1, e_3, e_1e_3/\tilde{e}_2 \gg E_j, \hat{E}_j \gg \sqrt{e_1e_3/\tilde{e}_2 m_0}$. 
 All of these are easily arranged.
\section{Non-supersymmetric vacua} \label{nonsusy}
In general, besides the supersymmetric vacua identified in the previous section, we expect there to be additional vacua that are non-supersymmetric. To identify these we should find the effective potential in the four-dimensional Einstein frame, in which the four-dimensional Einstein-Hilbert action has the standard normalisation. However, the axion fields $x_k$ and $X_j$, defined in (\ref{C3}), enter the ten-dimensional action (\ref{SIIA})
only via the $C_3 \wedge H_3^{bg}\wedge dC_3$ term in the Chern-Simons piece. This term
is only non-zero if $dC_3$ is ``polarised'' in the four-dimensional spacetime directions,
 {\it i.e.} $dC_3=fd^4x \equiv \mathcal{F}_0$; it has no physical degrees of freedom and can be treated as a
 Lagrange multiplier. The part of the action involving $\mathcal{F}_0$ has the form
 \beq
 S=-\frac{1}{2\kappa_{10}^2}\int (\mathcal{F}_0 \wedge ^*\!\!\mathcal{F}_0+2\mathcal{F}_0 \wedge X)
 \eeq
 where
 \beq
 X=F_6^{bg}+B_2 \wedge F_4^{bg}+C_3 \wedge H_3^{bg} -\frac{m_0}{6}B_2 \wedge B_2 \wedge B_2
 \eeq
 The equation of motion for $\mathcal{F}_0$ gives
 \beq
 ^*\!\mathcal{F}_0+X=0
 \eeq
 Then subsituting back gives
 \beq
 S=-\frac{1}{2\kappa_{10}^2}\int X \wedge ^*\!\!X \label{XX}
 \eeq
 which is stationary when $X=0$. The equation that stabilises the axion follows from
 \beq
 \int X=0=\int\left(F_6^{bg}+B_2 \wedge F_4^{bg}+C_3 \wedge H_3^{bg} -\frac{m_0}{6}B_2 \wedge B_2 \wedge B_2 \right) \label{XDW}
\eeq
Using (\ref{Jc}), (\ref{C3}), (\ref{H3bg}) and (\ref{F4bg}) this gives
\bea
\frac{8}{3}(x_0p_0-x_1p_1)-24 \sum _jX_jP_j&=&e_0-\frac{4}{3}\sum _jb_je_j+4\sum _j(B_jE_j+3\hat{B}_j\hat{E}_j)
+4b_2\sum_j(G_j+3 \hat{G}_j)         \nonumber \\
&-& \frac{4m_0}{3}b_1b_2b_3+2m_0b_2 \sum _j(B_j^2+3\hat{B}_j^2) \label{axion2}
\eea
This fixes the {\em same} linear combination of the axions $x_0, x_1$ and $X_j$ as in (\ref{axion}), and indeed, using (\ref{WK2}), the value agrees with that found in the supersymmetric treatment when the Kalb-Ramond fields $b_k, B_j$ and $\hat{B}_j$ have the values given in (\ref{B0}).

The remaining moduli  are stabilised by minimising the effective potential
 $V$  in the Einstein frame with metric $g_{\mu \nu}^E$.  We pass to this frame by redefining the four-dimensional metric
\beq
g_{\mu \nu} = \frac{e^{2\phi}}{{\rm Vol}(\mathcal{M})}g_{\mu \nu}^E \label{gE}
\eeq
where ${\rm Vol}(\mathcal{M})$ is the volume of the compact space $\mathcal{M}=T^6/\mathbb{Z}_6'$
\beq
{\rm Vol}(\mathcal{M})\equiv \int_{T^6/\mathbb{Z}_6'} d^6y\sqrt{g_6}
\eeq
with $g_6$ the determinant of the 6-dimensional metric.
Invariance of the 6-dimensional K\"ahler metric
under the action of the point group and the orientifold projection  $\mathcal {R}$ requires that
\beq
ds^2=\gamma _1 dz^1 d \bar{z}^1+\gamma _2 dz^2 d \bar{z}^2+\gamma _3 dz^3d\bar{z}^3
 \label{ds2}
\eeq
where the $\gamma _i \   (i=1,2,3)$ are real and positive.   In the $\theta^3$-twisted sector there are 16 $\mathbb{Z}_2$ fixed points   $f_{i,j} \in T^2_1 \times T^2_3$ with $i,j=1,4,5,6$, defined in (\ref{fij}) and (\ref{1456}). These fixed points are blown up using the Eguchi-Hanson $EH_2$ metric
\beq
ds^2=g_{k,\bar{\ell}}dz^kd\bar{z}^{\ell}
\eeq
where $k,\ell=1,3$ and
\beq
g_{k,\bar{\ell}}=\Gamma[A(u)\delta _{k\bar{\ell}}+B(u)(z_k-Z_k)(\bar{z}_{\ell}-\bar{Z}_{\ell})]
\eeq
when $f_{i, j}$ is at $(z^1,z^3)=(Z^1,Z^3) \in T^2_1 \times T^2_3$. The  functions $A(u)$ and $B(u)$ are
given by
\bea
A(u)&=&u^{-1}(\lambda^4+u^2)^{1/2}\lambda ^4 \\
B(u)&=&A'(u)
\eea
with $\lambda $ the blow-up parameter and $u$ as defined in (\ref{u}). In general, both the twisted modulus $\Gamma$ and the blow-up parameter $\lambda$  depend on the fixed point $f_{i,j}$ with which they are associated. However, the transformation property (\ref{e456j}) of the twisted 2-forms, or rather the analogous property of the twisted 2-cycles, shows that $\hat{\Gamma}_j$ and $\hat{\lambda}_j$, associated with $f_{4,j}, f_{5,j}$ and $f_{6,j}$, are independent of the $T^2_1$ fixed point $i=4,5,6$; the corresponding parameters for $f_{1,j}$ are denoted by $\Gamma _j$ and $\lambda _j$.
In the untwisted sector there are then three real  moduli and
\beq
{\rm Vol}(\mathcal{M})=\frac{1}{6}\prod _{k=1,2,3}{\rm Vol}(T^2_k)= \frac{1}{6}\gamma _1 \gamma _2 \gamma _3 {\rm Vol}_6
\label{VolM}
\eeq
where ${\rm Vol}_6$ is defined in (\ref{Vol6}) and (\ref{Volk}).
The (4-dimensional) volume of the blow-up is
\beq
{\rm Vol}(f_{i,j})=\Gamma ^2\frac{1}{4} \pi ^2\lambda^4
\eeq
taking $0 \leq u \lesssim \lambda^2$. The local analysis that we carry out here is valid provided that the volume of the blow-up modes is small compared with the untwisted volume ${\rm Vol}(T^2_1) {\rm Vol}(T^2_3)$ of the 4-torus containing them, {\it i.e.} provided that $\Gamma ^2 \pi ^2\lambda ^4 \ll {\rm Vol}(T^2_1) {\rm Vol}(T^2_3)$.
Blowing up $f_{i,j}$ in this manner removes a volume ${\rm Vol}(f_{i,j})$ from the untwisted volume ${\rm Vol}(T^2_1) {\rm Vol}(T^2_3)$. 
With $g_{\mu \nu}^E$ as given in (\ref{gE}), the effective potential $V$ is defined by
\beq
S= \frac{1}{\kappa _{10}^2}\int d^4x\sqrt{-\det( g^E)}(-V)
\eeq
Taking $F_2^{bg}=0$, as in (\ref{WK2}), there are four contributions to $V$
\beq
V=V_H+V_F+V_{m_0}+V_{BI}
\eeq
 deriving respectively from the $|H_3|^2$, $|F_4|^2$, $m_0^2$ and the Born-Infeld terms
 in (\ref{SIIA}). 
With $H_3^{bg}$ given by (\ref{H3bg}), we find
\beq
V_H=h\frac{e^{2\phi}}{{\rm Vol}^2(\mathcal{M})} \label{VH}
\eeq
where
\beq
h=\frac{2}{3}(p_0^2+p_1^2)+6\sum_jP_j^2 
\eeq
on both lattices. As noted previously, $h$ is fixed by the integers given in equations (\ref{p0p1a}) ... (\ref{Pjb}), independently of the coordinate scales $R_{1,3,5}$.
Similarly, with
$F_4^{bg}$ given by (\ref{F4bg}), we find
\bea
V_F&=&\frac{e^{4\phi}}{{\rm Vol}^3(\mathcal{M})}
\left( \frac{2}{9}\sum _{k=1,2,3}e_k^2{\rm Vol}(T^2_k)^2+16 \frac{{\rm Vol}(\mathcal{M})}{{\rm Vol}(T^2_2)}\sum_j (E_j^2+3\hat{E}_j^2)+ \right.\nonumber \\
&&+ \left.\frac{1}{6}{\rm Vol}(\mathcal{M}){\rm Vol}(T^2_2)\sum_j\left[\frac{G_j^2}{{\rm Vol}(f_{(1,j)})}    +
\frac{3\hat{G}_j^2}{{\rm Vol}(f_{(4,j)})}    \right]  \right)  \label{VF}
\eea
where
\beq
{\rm Vol}(T^2_k)=\gamma _k {\rm Vol}_k \quad  {\rm for} \quad k=1,2,3 
\eeq 
with ${\rm Vol}_k$ defined in (\ref{Volk}). 
Likewise
\beq
V_{m_0}=\frac{m_0^2e^{4\phi}}{2{\rm Vol}(\mathcal{M})}=\mu\frac{m_0^2e^{4\phi}}{{\rm Vol}(\mathcal{M})} \label{Vm0}
\eeq
with $\mu =1/2$.

As in \cite{DeWolfe:2005uu}, the only terms relevant to the stabilisation of the twisted moduli are $V_F$ and $V_{m_0}$, since the former dominates as  ${\rm Vol}(f_{i,j})\rightarrow 0$ and the latter as ${\rm Vol}(\mathcal{M})\rightarrow \infty$. In equation (\ref{Vm0}) we may write
\beq
{\rm Vol}(\mathcal{M})={\rm Vol}_0(\mathcal{M})-\frac{1}{6}{\rm Vol}(T^2_2)\sum_j [{\rm Vol}(f_{1,j})+3{\rm Vol}(f_{4,j})]
\eeq
where ${\rm Vol}_0(\mathcal{M})={\rm Vol}(T^2_1) {\rm Vol}(T^2_2){\rm Vol}(T^2_3)/6$ is the volume with no blow up. Then, minimising the potential gives
\bea
{\rm Vol}(f_{1,j})=\frac{|G_j|}{\sqrt{3} |m_0|}  \label{volf1j}\\
{\rm Vol}(f_{4,j})=\frac{|\hat{G}_j|}{\sqrt{3} |m_0|} \label{volf4j}
\eea
and we are justified in using this local treatment provided that the $F_4^{bg}$ fluxes are chosen so that
\beq 
|G_j,\hat{G}_j| \ll \sqrt{3}|m_0| {\rm Vol}(T^2_1) {\rm Vol}(T^2_3) \label{Gbound}
\eeq
With these values for the blow-up volume
\beq
V_F=V_{F1}+V_{F2}
\eeq
where
\bea
V_{F1}&=&\frac{e^{4\phi}}{{\rm Vol}^3(\mathcal{M})}
\left( \frac{2}{9}\sum _{k=1,2,3}e_k^2{\rm Vol}(T^2_k)^2+16 \frac{{\rm Vol}(\mathcal{M})}{{\rm Vol}(T^2_2)}\sum_j (E_j^2+3\hat{E}_j^2) \right) \\
V_{F2}&=& \frac{\sqrt{3}e^{4\phi}|m_0|{\rm Vol}(T^2_2)}{6{\rm Vol}^2(\mathcal{M})}
 \sum_j(|G_j|+3|\hat{G}_j|)   
\label{VF2}
\eea
The Born-Infeld term gives
\beq
V_{BI}=\mu_6 \kappa_{10}^2\frac{e^{3\phi}}{{\rm Vol}^2(\mathcal{M})}\sum_{\kappa}N_{\kappa}\int_{\kappa} d^3\xi\sqrt{\det(g_3)}
\eeq
and using the (bulk part of the) tadpole cancellation condition given in (\ref{RR}), we can rewrite this as
\beq
V_{BI}=-\frac{1}{4}\frac{e^{3\phi}}{{\rm Vol}^2(\mathcal{M})}\int_{\Pi_{m_0H_3^{bg}}} d^3\xi\sqrt{\det(g_3)} \label{VBI}
\eeq
where $\Pi_{m_0H_3^{bg}}$ is the 3-cycle of which the field $m_0H_3^{bg}$ is the Poincar\'e dual; $H_3^{bg}$ is given in (\ref{H3bg}). For the two cases of interest, as shown in \cite{Bailin:2008xx},
\bea
\Pi_{m_0H_3^{bg}}=\frac{\sqrt{{\rm Vol}_6}}{9R_1R_3R_5}m_0\left[(p_1-p_0) \rho _1-(p_0+p_1)(\rho _4+2\rho _6)\right] -\frac{  2i\sqrt{  {\rm Vol}_2  }   }{(1-2\alpha)R_3}\sum _jP_j(2\epsilon _j+\tilde{\epsilon}_j)   \label{pi3A} \\
=-\frac{  \sqrt{  {\rm Vol}_6  }    }{ 9\sqrt{3}R_1R_3R_5  } m_0\left[(p_0+p_1)\rho _6+(p_1-p_0)(\rho _3+2\rho _1) \right]  -\frac{  2i\sqrt{  {\rm Vol}_2  }   }{(1-2\alpha)R_3}\sum _jP_j(\epsilon _j+2\tilde{\epsilon}_j)          \label{pi3B}
\eea
 for {\bf AAA} and {\bf BAA} respectively. To calculate
 the integral in (\ref{VBI}), we use the result \cite{Becker:1995kb} quoted in \cite{DeWolfe:2005uu}, since $\Pi _{m_0H_3^{bg}}$ is a special Lagrangian 3-cycle. 
The holomorphic 3-form 
 $\Omega$, defined in (\ref{Oma}), is normalised by demanding that
\bea
i\int _{\mathcal{M}}\Omega \wedge \bar{\Omega} =1&=&\frac{16}{3}(Z_1g_1-Z_0g_0)+48 \sum _jY_jf_j
 \label{Z1Z0} \\
&\equiv& \frac{32}{3}\mathcal{G}(Z_0,Z_1,Y_j) \label{G}
\eea
Then, according to the calibration formula
\beq
\int_{\Pi_{m_0H_3^{bg}}} d^3\xi\sqrt{\det(g_3)}=\sqrt{2{\rm Vol}(\mathcal{M})}\int_{\Pi_{m_0H_3^{bg}}}(\Omega+\bar{\Omega})
\eeq
So
\beq
V_{BI}=-b|m_0|\frac{e^{3\phi}}{{\rm Vol}^{3/2}(\mathcal{M})}
\eeq
where
\beq
b=2\sqrt{2}\left|\frac{2}{3}(p_1Z_1-p_0Z_0)+6\sum _jY_jP_j \right| \label{b}
\eeq
for both lattices.

The various contributions to $V$ are homogeneous in ${\rm Vol}(T^2_k)$. Hence at the stationary point
\bea
0=\sum_k{\rm Vol}(T^2_k)\frac{\partial V}{\partial {\rm Vol}(T^2_k)}&=&6V_H+7V_{F1}+5V_{F2}+3V_{m_0}+\frac{9}{2}V_{BI} \\
0=\frac{\partial V}{\partial \phi}&=&2V_H+4V_{F1}+4V_{F2}+4V_{m_0}+3V_{BI} 
\eea
Eliminating $V_{F1}$ gives
\beq
10V_H=8V_{F2}+16V_{m_0}+3V_{BI} \label{VHmBI}
\eeq
Also, we require that $\partial V/\partial {\rm Vol}(T^2_k) =0$, which gives
\bea
&&|e_1{\rm Vol}(T^2_1)|^2 =|e_3{\rm Vol}(T^2_3)|^2\equiv y^2  =\nonumber\\
&&=|e_2{\rm Vol}(T^2_2)|^2-6{\rm Vol}(T^2_1){\rm Vol}(T^2_3) \sum_j(E_j^2+3\hat{E}_j^2)+\frac{9\sqrt{3}|m_0|{\rm Vol}(\mathcal{M})^2}{4{\rm Vol}(T^2_1){\rm Vol}(T^2_3)}\sum _j(|G_j|+3 |\hat{G}_j|) \nonumber \\
&& \label{ydef} 
\eea
(with $y>0$).
It follows that
\beq
|e_2{\rm Vol}(T^2_2)|^2=y^2(1+\epsilon)-\frac{\eta}{y^2}{\rm Vol}(\mathcal{M})^2 \label{e2VolT2}
\eeq
where 
\bea
\epsilon &\equiv& \frac{6}{|e_1e_3|}\sum_j (E_j^2+3\hat{E}_j^2)  \label{epsdef}\\
\eta &\equiv & \frac{9\sqrt{3}|m_0e_1e_3|}{4}\sum_j(|G_j|+3|\hat{G}_j|) \label{etadef}
\eea
 The requirement (\ref{Gbound}) that justifies the local treatment gives
\beq
|G_j, \hat{G}_j| \ll \frac{\sqrt{3}|m_0|y^2}{|e_1e_3|}
\eeq
so that 
\beq
\eta \ll 27 (m_0y)^2
\eeq
We may also write ${\rm Vol}(\mathcal{M})$ in terms of $y$:
\beq
{\rm Vol}(\mathcal{M})=\frac{1}{6}\prod _k {\rm Vol}(T^2_k)=\frac{y^3(1 +\epsilon)^{1/2}}{(36|e_1e_2e_3|^2+\eta y^2)^{1/2}} \label{VMy}
\eeq
so that
\beq
|e_2{\rm Vol}(T^2_2)|^2=\frac{y^2(1+\epsilon)}{1+\frac{\eta y^2}{36|e_1e_2e_3|^2} }
\eeq  
Defining
\beq
x \equiv e^{\phi} \sqrt{{\rm Vol}(\mathcal{M})} \label{xdef}
\eeq
it follows from (\ref{VHmBI}) that
\beq
10h=\left(16 \mu  m_0^2 +\frac{32 \eta}{9y^2}\right)x^2-3b |m_0|x \label{x}
\eeq
 which fixes $x$ as a function of $y$. Hence
\beq
|m_0|x=\frac{3b}{32(\mu+\frac{2\eta}{9m_0^2y^2})}\left(1+\sqrt{1+\frac{640(\mu+\frac{2\eta}{9m_0^2y^2}) h}{9b^2}}\right) \label{m0x}
\eeq
and at the stationary point, we may eliminate the dilaton and express the potential in terms of $y$ alone:
\beq
V=\frac{A}{{\rm Vol}(\mathcal{M})^3}+\frac{B}{{\rm Vol}(\mathcal{M})^5}
\eeq
where ${\rm Vol}(\mathcal{M})$ is given by (\ref{VMy}) and
\bea
A&\equiv& hx^2+\mu m_0^2 x^4 -b|m_0|x^3  \label{defA}\\
B &\equiv& \frac{2}{3}x^4y^2(1+\epsilon)\left(1+\frac{\eta y^2}{3(36|e_1e_2e_3|^2+\eta y^2)}\right) \label{volm5}
\eea
with $x$ given by (\ref{x}). 
Since $B>0$, it is easy to see that the potential $V \rightarrow +\infty$ as $y \rightarrow 0+$. 
Similarly, $V \rightarrow 0$ as $y \rightarrow \infty$. The limit is approached from above or below depending upon the sign of $A$ in this region. If $A<0$, then there is certainly an anti-de-Sitter minimum at a finite value of $y$; otherwise, no conclusion can be reached without a  more detailed consideration of the parameters.  It follows from (\ref{defA}) and (\ref{x}) that
\beq
A\sim \frac{13}{3}x^2( h-\mu m_0^2x^2) \quad {\rm as} \quad y \rightarrow \infty
\eeq
In the same limit, (\ref{m0x}) gives
\beq
|m_0|x \simeq \frac{3b}{32\mu}\left(1+\sqrt{1+\frac{640\mu}{9b^2}} \right)
\eeq
Then $A<0$   if and only if  
\beq
b^2>4\mu h \label{b2muh}
\eeq

To proceed further, we need to know the dependence of the moduli $g_{0,1},f_j$ that appear in (\ref{Z1Z0}) on $Z_{0,1},Y_j$. For simplicity, we consider only the bulk contributions $g_{0,1}$ and assume that these derive from a homegeneous quadratic prepotential $\mathcal{G}$, defined in (\ref{G}), of the form 
\beq
\mathcal{G}(Z_0,Z_1) = \alpha Z_0^2+2\beta Z_0Z_1+\gamma Z_1^2 
\eeq
with $\alpha, \beta$ and $\gamma $ (real) constants (not functions of $Z_0/Z_1$). 
Then the moduli $g_{0,1}$ are given by
\bea
g_0&=&-\frac{\partial \mathcal{G}}{\partial Z_0}=-2(\alpha Z_0+\beta Z_1) \\
g_1&=&\frac{\partial \mathcal{G}}{\partial Z_1}=2(\beta Z_0+\gamma Z_1)
\eea
The question we address is whether $\mathcal{G}$ may be chosen so that (\ref{b2muh}) is always satisfied. Keeping only the bulk contributions, the minimum value of 
\beq
b^2=\frac{32}{9}(Z_0p_0-Z_1p_1)^2
\eeq
subject to the constraint (\ref{Z1Z0}) that $\mathcal{G}(Z_0,Z_1)=3/32$ is
\beq
b^2=3\frac{\gamma p_0^2+2\beta p_0p_1+\alpha p_1^2}{\alpha \gamma -\beta ^2}
\eeq
Evidently, we may ensure that (\ref{b2muh}) is satisfied by choosing $\alpha, \beta, \gamma$ sufficiently small. On the {\bf AAA} lattice, 
\beq
\frac{p_0}{p_1}=3+2t^c_1=5,1
\eeq
so that the minimum value of $b^2$  and $2h$ are
\bea
b^2&=&\frac{3}{p_1^2(\alpha \gamma -\beta ^2)}[15\gamma+6\beta+\alpha+4t^c_1(3\gamma+\beta)] \\
&=&\frac{3}{p_1^2(\alpha \gamma -\beta ^2)}(25\gamma+10\beta+\alpha, \gamma+2\beta+\alpha) \\
2h&=&\frac{8}{3p_1^2}(7+6t^c_1) \\
&=&\frac{8}{3p_1^2}(13,1)
\eea
for $t^c_1=+1,-1$ respectively. On the {\bf BAA} lattice, since $p_0=p_1$ in this case, $b^2$ and $2h$ have the same values as in the $t^c_1=-1$ case for the {\bf AAA} lattice.


The untwisted part of the 4-form flux $F_4^{bg}$ given in equation (\ref{F4bg}). It is specified by the quantities $e_k {\rm Vol}_k/{\rm Vol}_6  \ (k=1,2,3)$. Using (\ref{ydef}), the ratios of the metric moduli $\gamma _i/\gamma _j=e_j{\rm Vol}_j/e_i {\rm Vol}_i$ are specified for a given value of $F_4^{bg}$. The minimisation of $V$ fixes $y^2/|e_1e_2e_3|^2$, and $F_4^{bg}$ also specifies the combination $|e_1e_2e_3|/{\rm Vol}_6^2$. Thus, the stabilisation fixes the overall scale of the metric moduli $(\gamma_1\gamma _2\gamma _3)^2=(y^2/ |e_1e_2e_3|)^3(|e_1e_2e_3|/{\rm Vol}_6^2)$ in terms of the specified background fluxes. Similarly, 
the (untwisted) background flux $H_3^{bg}$, defined in equation (\ref{H3bg}), is specified by $p_{0,1}/\sqrt{{\rm Vol}_6}$. Thus equation (\ref{m0x})  fixes $x/\sqrt{{\rm Vol}_6}$ in terms of the background fluxes  $m_0$ and  $H_3^{bg}$.  With $x$ defined in (\ref{xdef}), it follows that $x/\sqrt{{\rm Vol}_6} \simeq e^{\phi} \sqrt{\gamma _1 \gamma_2 \gamma _3}$, and since the moduli $\gamma _{1,2,3}$ have already been fixed,  this result stabilises the dilaton $\phi$ in terms of the background fluxes. The argument may be extended to include the twisted moduli.
\section{Stability} \label{stab}
Since we have taken $F_2^{bg}=0$, the $|F_2|^2$ and $|F_4|^2$ terms in the the action $S_{IIA}$, given in (\ref{SIIA}), are at least quadratic in the fields $B_2$, there being no $\mathbb{Z}_6'$-invariant 1-form fields $C_1$. The Chern-Simons terms have already been accounted for in the minimisation of $X$. Thus the whole action $S_{IIA}$ is at least quadratic  in the moduli fields $b_k, \ B_j, \ \hat{B}_j$ defined in (\ref{tk}) ... (\ref{hatTj}) and (\ref{tknew}) ... (\ref{hatTjnew}), and we may consistently set all of their expectation values to be zero, as in (\ref{B0}) in the supersymmetric case.
However, there are  fluctuations  $b_{k}(x), B_j(x), \hat{B}_j(x)$ around this
 solution, and
the $B_2 \wedge B_2 \wedge ^*\!\!F_4^{bg}$ contribution to $|F_4|^2$ can make the solution unstable if the mass matrix for the fluctuations  has a negative eigenvalue.

After eliminating the Lagrange multiplier $\mathcal{F}_0 \equiv dC_3$, the effective action  deriving from this field is given in (\ref{XX}) with $X$ in (\ref{XDW}).
With the $B_2$-moduli set to zero, the stabilised linear combination of the axions  given in (\ref{axion}) reduces to
\beq
\frac{8}{3}(x_0p_0-x_1p_1)-24 \sum _jX_jP_j=e_0
\eeq
 The $B_2 \wedge F_4^{bg}+C_3 \wedge H_3^{bg}$ piece in $X$ is linear in the fluctuation fields and the above stabilised combination of  axion fields.  Hence the action (\ref{XX}) mixes them and we need to consider the  quadratic terms, including kinetic terms,  for both sets of fields simultaneously. The unstabilised (orthogonal) axion fields  are, of course, massless.

 The kinetic terms for the $B_2$ field fluctuations derive from the contribution
\bea
-\frac{1}{2\kappa_{10}^2}\int d^{ 10}x \ \sqrt{-g}\frac{1}{2} \ e^{-2\phi}|H_3|^2 &\supset& -\frac{1}{2\kappa_{10}^2}\int d^{ 10}x \ \sqrt{-g}e^{-2\phi}\frac{1}{2}dB_2 \wedge ^*\!\!dB_2 \\
&=& -\frac{1}{2\kappa_{10}^2}\int d^{ 4}x \ \sqrt{-g_E}\mathcal{L}_K(B)
\eea
where the kinetic Lagrangian density is
\beq
\mathcal{L}_K(B)=\frac{1}{2}\sum_{k=1,2,3}(\partial _{\mu} \tilde{b}_k)(\partial ^{\mu} \tilde{b}_k) +
\frac{1}{2}\sum_{j=1,4,5,6}[(\partial _{\mu} \tilde{B}_j)(\partial ^{\mu} \tilde{B}_j) + (\partial _{\mu} \tilde{\hat{B}}_j)(\partial ^{\mu} \tilde{\hat{B}}_j)]
\eeq
with $\partial ^{\mu} \tilde{b}_k=g^{\mu \nu}_E\partial _{\nu} \tilde{b}_k$ {\it etc.},
and the fields $\tilde{b}_k,\tilde{B}_j, \tilde{\hat{B}_j}$  defined so that they are canonically normalised:
\bea
\tilde{b}_{k} &\equiv& \frac{2b_k}{{\rm Vol}(T^2_k)}  \label{bktil}\\
\tilde{B}_j &\equiv& \sqrt{\frac{2{\rm Vol}(T^2_2)}{ {\rm Vol}(\mathcal{M})}}B_j \label{Bjtil}\\
\tilde{\hat{B}}_j &\equiv& \sqrt{\frac{6{\rm Vol}(T^2_2)}{ {\rm Vol}(\mathcal{M})}}\hat{B}_j \label{tilhatBj}
\eea
Quadratic terms  in  these fields arise from
\bea
-\frac{1}{2\kappa_{10}^2}\int d^{ 10}x \ \sqrt{-g}[|F_2|^2+|F_4|^2] &\supset&\!\!\!\! -\frac{1}{2\kappa_{10}^2}\int  [m_0^2B_2 \wedge ^*\!\!B_2-m_0B_2 \wedge B_2 \wedge ^*\!\!F_4^{bg}] \nonumber \\
=-\frac{1}{2\kappa_{10}^2}\int d^4x \sqrt{-g_E} \ \frac{e^{4\phi}}{{\rm Vol}^2(\mathcal{M})}\!\!\!&& \!\!\!\!\left(\sum _k\left[ m_0^2{\rm Vol}(\mathcal{M})\tilde{b}_k\tilde{b}_k  + \frac{4}{3}m_0\tilde{b}_1\tilde{b}_2\tilde{b}_3\frac{e_k{\rm Vol}(T^2_k)}{\tilde{b}_k} \right] \right. \nonumber \\
+\left. \sum _j\left[m_0^2{\rm Vol}(\mathcal{M})-\frac{2}{3}m_0 e_2 {\rm Vol}(T^2_2) \right](\tilde{B}_j ^2 +\tilde{\hat{B}}_j^2)\right.&+&\!\!\!\!16m_0\left.
\sqrt{\frac{{2 } {\rm Vol}(\mathcal{M})}{{\rm Vol}(T^2_2)} }\tilde{b}_2\sum _j (\tilde{B}_jE_j+\sqrt{3}\tilde{\hat{B}}_j\hat{E}_j) \right. \nonumber \\
\left.
-4m_0{\rm Vol}(T^2_2)\tilde{b}_1\tilde{b}_3  \sum_j(G_j+3 \hat{G}_j)\right. &-& \left.
  \!\!\!\!  \frac{2m_0^2{\rm Vol}(\mathcal{M})}{ \sqrt{3}}\sum _j\left[\tilde{B}_j^2 s_j+   \hat{\tilde{B}}_j^2\hat{s}_j  \right] 
\right) 
\eea
where $s_j, \hat{s}_j$ are respectively the signs of $G_j/m_0, \hat{G}_j/m_0$, and the last term follows when we substitute the stabilised values (\ref{volf1j}) and (\ref{volf4j}) of the blow-up volumes.

The kinetic terms for the $C_3$ fluctuations  arise from
\bea
-\frac{1}{2\kappa_{10}^2}\int d^{ 10}x \ \sqrt{-g}|F_4|^2 &\supset& -\frac{1}{2\kappa_{10}^2}\int dC_3 \wedge ^*\!\!dC_3 \\
=-\frac{1}{2\kappa_{10}^2}\int d^4x \ \sqrt{-g_E} \ \frac{e^{2\phi}}{{\rm Vol}(\mathcal{M})}
\!\! && \!\!\!\!\left[\frac{4}{3}
(\partial (x_0-x_1))^2
+\frac{4}{3}(\partial(x_0+x_1))^2+6\sum _j(\partial X_j)^2\right] \nonumber \\
&&\\
= -\frac{1}{2\kappa_{10}^2}\int d^{ 4}x \ \sqrt{-g_E}\!\!\!  &&\!\!\! \frac{1}{2}\left[(\partial \tilde{x}_1)^2+(\partial \tilde{x}_2)^2+\sum _j(\partial \tilde {X}_j)^2 \right]
\eea
 and the canonically normalised fields $\tilde{x}_{1,2}$ and $\tilde{X}_j$ are  given by
\bea
\tilde{x}_1 &\equiv&     \sqrt{\frac{2}{3    {\rm Vol}(\mathcal{M})   }} \ 2e^{\phi}(x_0-x_1) \label{tilx1} \\
\tilde{x}_2 &\equiv & \sqrt{\frac{2       }{3    {\rm Vol}(\mathcal{M})   }} \ 2e^{\phi}(x_0+x_1) \label{tilx2}\\
\tilde{X}_j &\equiv& \sqrt{\frac{3}{{\rm Vol}(\mathcal{M})}} \ 2e^{\phi}X_j \label{tilXj}
\eea
Quadratic terms in  $b_k$ and $x_{0,1}$ arise from (\ref{XX})
\bea
 S=-\frac{1}{2\kappa_{10}^2}\int X \wedge ^*\!\!X \supset -\frac{1}{2\kappa_{10}^2}\int (B_2 \wedge F_4^{bg}+C_3 \wedge H_3^{bg}) \wedge ^*\!\!(B_2 \wedge F_4^{bg}+C_3 \wedge H_3^{bg}) \label{S2bx}
 \eea
As noted previously, the only coupled combination of axion fields corresponds to the stabilised axion, whose normalised field $\tilde{a}$ is given in terms of the rescaled fields  $\tilde{x}_{1,2},\tilde{X}_j$ by
\bea
p_0 x _0-p_1x_1-9\sum_jP_jX_j \propto  (p_0+p_1)\tilde{x}_1+(p_0-p_1) \tilde{x}_2 -6\sqrt{2}\sum _j P_j\tilde{X}_j \equiv N\tilde{a} \label{Na}
\eea
where 
\beq
N=\sqrt{2}\left[p_0^2+p_1^2+36\sum _j P_j^2\right]^{1/2}
\eeq
 We shall consider only the {\em untwisted} contibutions.
 Then the quadratic terms deriving from (\ref{S2bx}) are
\beq
S=-\frac{1}{2\kappa_{10}^2}\int d^4x \sqrt{-g_E} \ \frac{4e^{4\phi}}{9{\rm Vol}^3(\mathcal{M})}\left( \sum _k {\rm Vol}(T^2_k)\tilde{b}_k e_k+{e^{-\phi}}\sqrt{3 {\rm Vol}(\mathcal{M})(p_0^2+p_1^2)}\tilde{a}\right)^2 \label{SBFCH}
\eeq
Stability requires that the eigenvalues of the mass matrix are all positive. However the uncoupled axion is massless, so the best we can hope for is that the remaining four mass eigenstates are non-tachyonic.
The mass matrix deriving from (\ref{SBFCH}) may be written in the form
\bea
{\bf m}^2&=&\frac{2 e^{4\phi}m_0^2 }{ {\rm Vol}(\mathcal{M})}\left( \begin{array}{cccc}
1+\alpha^2\gamma^2& \alpha \gamma(s_3 +s_1s_2\alpha)&\alpha(s_2 +s_3s_1\alpha\gamma^2)&s_1\alpha \beta \gamma\\

\alpha \gamma(s_3 +s_1s_2\alpha)&1+\alpha^2 & \alpha \gamma(s_1+s_2s_3\alpha )&s_2\alpha \beta \\

\alpha(s_2 +s_3s_1\alpha\gamma^2)& \alpha \gamma(s_1+s_2s_3\alpha )&1+\alpha^2\gamma^2 &s_3\alpha \beta \gamma\\

s_1\alpha \beta \gamma&s_2\alpha \beta &s_3\alpha \beta \gamma & \beta^2

\end{array}
\right) \nonumber \\
&&
\eea
where
\bea
\alpha &\equiv& \frac{4|e_1e_2e_3|}{|m_0|y^2} \\
\beta &\equiv &\frac{2(p_0^2+p_1^2)^{1/2}}{\sqrt{3}m_0x} \\
\gamma &\equiv& \left(1+\frac{\eta y^2}{36 |e_1e_2e_3|^2}\right)^{1/2}(1+\epsilon)^{-1/2} 
\eea
and $s_{1,2,3}=\pm 1$ are the signs of $e_{1,2,3}$. 
The general expressions for the eigenvalues are too large to be tractable, but positive-definiteness is ensured provided that the following quantities are all positive:
\bea
{\rm tr}({\bf m}^2)&=&\sum_i (m^2)_i=\beta^2+3+a^2(1+2\gamma^2) \\
\det({\bf m}^2)=\prod_i (m^2)_i&=&\beta^2(1-a^2+2a^3\gamma^2-2a^2\gamma^2) \equiv d(a)\\
\sum_{i,j} (m^2)_i(m^2)_j&=&3\beta^2+6a^3\gamma^2+2a^2\gamma^2+3+a^2  \equiv d_4(a)\\
\sum_{i,j,k} (m^2)_i(m^2)_j(m^2)_k&=&3\beta^2+1+4a^4\gamma^2+4a^3\gamma^2-a^4-2a^2\beta^2\gamma^2-a^2\beta^2 \equiv d_6(a)
\eea
where
\beq
a \equiv s_1s_2s_3 \alpha
\eeq
When $\gamma ^2 >1/4$ it is obvious that for large, positive values of $a \gg 1$  all of these {\em are} positive. The question is  whether there are other values, in particular negative values, for which we also have positive-definiteness, and what can be said when $\gamma ^2 \leq 1/4$. By inspection it is clear that the trace is automatically positive.
 For  general (non-zero) values of $\beta$ and $\gamma$, 
$\det{({\bf m}^2)}=d(a)>0$ provided that
\bea
&&a_1 \equiv \frac{1}{4\gamma^2}\left(1-\sqrt{1+8\gamma^2}\right)<a< a_2 \label{x2} \\
 {\rm or} && a_3<a  \label{x1}
\eea
where 
\bea
a_2 =\frac{1}{4\gamma^2}\left(1+\sqrt{1+8\gamma^2}\right), \quad  a_3=1 \quad {\rm for}  \quad \gamma^2 >1  \label{gamlarge} \\
a_2=1, \quad a_3 =\frac{1}{4\gamma^2}\left(1+\sqrt{1+8\gamma^2}\right) \quad  {\rm for}  \quad \gamma^2 <1 \label{gamsmall}
\eea
Note that $a_1$ is always negative, and $a_{2,3}$ positive. In the special case that $\gamma=0$, the function $d(a)=\beta^2(1-a^2)$ is positive only in the range $-1<a<1$.

We also require that $d_4(a)$ is positive. Evidently this is always the case for $a>0$, so  we need only consider whether negative values of $a$ lead to stronger constraints than those already derived. According to (\ref{x2}), the most negative value that we need to consider is $a=a_1$, which satisfies $d(a_1)=0$. For this value of $a$ it follows that 
\beq
d_4(a_1)=3 \beta^2+4a_1^2(1+2\gamma ^2)>0
\eeq
Further, $d_4(a)$ has only a single real (negative) root, so  the positivity of $\sum_{i,j} (m^2)_i(m^2)_j$ gives no further constraints.

 Finally, we require  also that $d_6(a)$ is positive. It is convenient to write
\beq
d_6(a)=N(a)-\beta ^2 D(a)
\eeq
where
\bea
N(a) &\equiv& (4\gamma ^2-1) a^4+4\gamma ^2 a^3+1  \label{Ndef}\\
&=&(a+1)[(4\gamma ^2-1) a^3+ a^2-a+1] \label{N2}\\
D(a)& \equiv& (2 \gamma ^2+1) a^2-3 \label{Ddef}
\eea
The special case in which $\gamma =0$ is easy to analyse. In this case $N(a)=1-a^4$ is positive only in the range $-1<a<1$ in which $d(a)$ is also positive. Since $D(a)=a^2-3$ is negative throughout this range, it is only for values of $a$ in this range that we have positive definiteness. 
The  case in which $\gamma ^2=1/4$ is also easy to analyse. Positivity of $d(a)$ requires that either $1-\sqrt{3}=a_1<a <a_2=1$ or $a>1+\sqrt{3}$. The function $N(a)=1+a^3$ is positive only when $a>-1$. Thus $N(a)$ is positive in both of these ranges, while  $D(a)=3(a^2-2)/2$ is negative in the region $a_1<a<a_2$, but positive in $a>a_3$. It follows that ${\bf m^2}$ is positive definite for any value of $\beta^2$ when  $a$ is in the range $a_1<a<a_2$, but only for values of $\beta^2<N(a)/D(a)$ in the range $a>a_3$.

The full analysis of the  conditions in which $d_6(a)$  and $d(a)$ are both positive for general values of $\gamma^2$  is given in Appendix B. The conclusions are as follows:
For values of $a$ in the range $a_1 <a <a_2$, the mass matrix ${\bf m}^2$ is  positive definite for all values of $\gamma^2$ and all values of $\beta^2$. If $0.1955 \lesssim \gamma ^2 <1/4$, there is in addition a finite region $a_3 < a <a_5$  in which ${\bf m}^2$ is positive definite but only  for values of $\beta^2$ that are bounded above by $N(a)/D(a)$. Finally, if $\gamma ^2 >1/4$, there is an infinite region $a>a_3$ in which ${\bf m}^2$ is positive definite, again for values of $\beta^2$ that are bounded above by $N(a)/D(a)$. Here $a_{1,2,3}$, defined in equations (\ref{x2}), (\ref{gamlarge}) and (\ref{gamsmall}), specify the regions in which $d(a)>0$, and $a_5$ is the  root  of the cubic factor in equation (\ref{N2}).

Although we have been discussing the conditions under which the (untwisted) mass eigenstates are non-tachyonic, in principle this is too strong a requirement in the anti-de Sitter space of our vacuum solutions. Tachyonic mass eigenstates are stable provided that they satisfy the Breitenlohner-Freedman bound \cite{Breitenlohner:1982bm, Breitenlohner:1982jf} 
\beq
m_i^2 \geq m_{BF}^2 \equiv -\frac{3}{4}|V_{\rm min}|
\eeq
where $-|V_{\rm min}|$ is the value of the potential at the anti-de Sitter minimum. The massless uncoupled axion obviously satisfies the bound, so it will not generate instability. However, determining which values of $a$ lead to other mass eigenstates that satisfy this weaker constraint is something that can only be done when $V_{\rm min}$ has actually been calculated, and this in turn requires a detailed consideration of the parameters, as already noted. The expectation or, more accurately, the hope is that when the anti-de Sitter minimum is lifted to Minkowski, in the manner of KKLT \cite{Kachru:2003aw}, then the tachyonic states will be lifted too. However, as Conlon has noted \cite{Conlon:2006tq}, it is not clear that {\em all} tachyons will be lifted by this mechanism. The uplifting is generally rather poorly controlled, and it is at least plausible that there may remain tachyons in the Minkowski space. 
\section{E2-instantons and Yukawa couplings} \label{se2}
 We have so far fixed only {\em one} linear combination of the axion fields. As noted previously, we may use non-perturbative  effects to stabilise the remaining axions. 
The non-perturbative effects under discussion are D$p$-branes that wrap non-trivial cycles in the compactification space $\mathcal{M}_6$, and that are pointlike in $\mathcal{M}_4$. Their world-volume is $(p+1)$-dimensional and { spacelike}, so they are Euclidean D$p$-branes, called E$p$-branes or E$p$-instantons for short. In type IIA string theory, $p$ is even and $p+1 \leq 6$. Hence $p=0,2,4$. Since there are no non-trivial 1- and 5-cycles on the orientifold $T^6/\mathbb{Z}_6'$ with which we are concerned, only E2-branes are relevant. 
The instanton action $S_{\rm inst}$  is given by \cite{Blumenhagen:2006xt}
\beq
S_{\rm inst}=2\pi \left(\frac{1}{g_s} \int_{\Xi}{\rm Re}  \ \Omega _3-i\int_{\Xi}C_3\right)  \label{SE2}
\eeq
where $\Xi$ is the 3-cycle wrapped by the E2-brane, and $\Omega _3$ is the holomorphic 3-form. Evidently an E2-instanton is coupled to the axion fields in $C_3$ and can lift (some of) the degeneracy of the axions that are not stabilised by the background flux $m_0H_3^{bg}$. 
Quite generally, we may expand $\Xi$ in terms of the untwisted 3-cycles $\rho _p \ (p=1,3,4,6)$ and the exceptional 3-cycles $\epsilon _j, \tilde{\epsilon}_j \ (j=1,4,5,6)$, so that
\beq
\Xi= \frac{1}{2}\sum _p Z_p\rho_p+ \frac{1}{2} \sum _j(z_j \epsilon _j+\tilde{z}_j \tilde{\epsilon}_j) \label{Xigen}
\eeq
where  $Z_p, z_j, \tilde{z}_j$ are integers. Supersymmetry constrains these coefficients. 
On the {\bf AAA} lattice, it requires that
\bea
3Z_3-2Z_4+Z_6=0 \label{susyY}\\
2Z_1-Z_3+Z_6>0 \label{susyX}
\eea
As displayed in equations (189) ... (194) of reference \cite{Bailin:2008xx}, there are three types of supersymmetric 3-cycle:
\bea
Z_p&=&(1,0,0,0) \bmod 2 \quad (n_k,m_k)= (1,0;1,0;1,0) \bmod 2\\
&=&(1,\theta,1,\theta) \bmod 2 \quad (n_k,m_k)= (1,1;\theta,1;1,1) \bmod 2  \\
&=&(0,0,1,0) \bmod 2 \quad (n_k,m_k)= (0,1;1,1;0,1) \bmod 2
\eea
(with $\theta=0,1$) called  respectively $c$-, $d_{\theta}$- and $e$-type. They are associated with exceptional parts having
\bea
c: \quad (z_{1,4}; \tilde{z}_{1,4}) \ {\rm or} \ (z_{5,6}; \tilde{z}_{5,6})= (0,0; 1,1) \bmod 2 \\
d_{\theta}: \quad (z_{1,6}; \tilde{z}_{1,6}) \ {\rm or} \ (z_{4,5}; \tilde{z}_{4,5})= (\theta, \theta; 1,1) \bmod 2 \label{dtheta} \\
e: \quad (z_{1,5}; \tilde{z}_{1,5}) \ {\rm or} \ (z_{4,6}; \tilde{z}_{4,6})= (0,0;1,1) \bmod 2 
\eea
 With the general form of the instanton's 3-cycle $\Xi$ given in equation (\ref{Xigen}), and with  $C_3$ on the {\bf AAA} lattice given in equation (\ref{C3}), we find
\bea
{\rm Im} \ S_{\rm inst}&=&-6\pi \left( \frac{2}{3^{1/4}}[(2Z_1-Z_3)(x_0+x_1)+Z_6(x_0-x_1)]+\frac{3^{1/4}}{\sqrt{2}}\sum _jX_jz_j \right) \\
& \propto & 2[(2Z_1-Z_3)r^{1/4}\tilde{x}_2+Z_6r^{-1/4}\tilde{x}_1]+\frac{1}{\sqrt{3}}\sum _j\tilde{X}_jz_j
\eea
using the canonically normalised fields defined in equations (\ref{tilx1}), (\ref{tilx2}) and (\ref{tilXj}). In general, this is
quite different from the combination $a$ given in equation (\ref{Na}) that is stabilised by the background flux.
Evidently a separate instanton is required for each of the unstabilised axions.

Similarly, on the {\bf BAA} lattice, supersymmetry requires that
\bea
2Z_3-Z_1-3Z_4=0 \label{susyYb}\\
Z_1-Z_4+2Z_6>0 \label{susyXb}
\eea
and, as displayed in equations (275) ... (280) of reference \cite{Bailin:2008xx}, again there are three types of supersymmetric 3-cycle:
\bea
Z_p&=&(0,1,0,0) \bmod 2 \quad (n_k,m_k)= (1,0;1,1;1,0) \bmod 2\\
&=&(\theta,1,\theta,1) \bmod 2 \quad (n_k,m_k)= (1,1;1,\theta;1,1) \bmod 2\\
&=&(0,0,0,1) \bmod 2 \quad (n_k,m_k)= (0,1;0,1;0,1) \bmod 2
\eea
(with $\theta=0,1$) called  respectively $c$-, $d_{\theta}$- and $e$-type. They are associated with exceptional parts having
\bea
c: \quad (z_{1,4}; \tilde{z}_{1,4}) \ {\rm or} \ (z_{5,6}; \tilde{z}_{5,6})= (1,1; 0,0) \bmod 2 \\
d_{\theta}: \quad (z_{1,6}; \tilde{z}_{1,6}) \ {\rm or} \ (z_{4,5}; \tilde{z}_{4,5})= (1,1:\theta, \theta) \bmod 2 \\
e: \quad (z_{1,5}; \tilde{z}_{1,5}) \ {\rm or} \ (z_{4,6}; \tilde{z}_{4,6})= (1,1; 0,0) \bmod 2 
\eea
In this case, we find that 
\bea
{\rm Im} \ S_{\rm inst}&=&-6\pi R_3\left( \frac{2}{3^{1/4}}[Z_1(x_0+x_1)-(Z_4-2Z_6)Z_6(x_0-x_1)]-\frac{3^{1/4}}{\sqrt{2}}\sum _jX_j\tilde{z}_j \right) \\
& \propto & 2[Z_1r^{1/4}\tilde{x}_2-(Z_4-2Z_6)r^{-1/4}\tilde{x}_1]-\frac{1}{\sqrt{3}}\sum _j\tilde{X}_j\tilde{z}_j
\eea
Again, this is generally quite different from the combination $a$ given in equation (\ref{Na}).

As noted in the Introduction, the surviving global $U(1)$ symmetries in our models forbid some of the Yukawa couplings that are needed to give non-zero masses to the quarks and leptons via the Higgs mechanism. Consider, for example, the model described by the fourth solution in Table 1 of reference \cite{Bailin:2008xx}. The weak hypercharge $Y$ is a linear combination of the $U(1)$ charges $Q_{a,c,d}$ associated respectively with the $SU(3)_c$ stack $a$, and the $U(1)$ stacks $c,d$. 
\beq
Y=\frac{1}{6}Q_a+y_cQ_c+\frac{1}{2}Q_d \label{Y}
\eeq
where $y_c=\pm \frac{1}{2}$.  
Using equations (63) and (66) of that paper,  the intersection numbers of $a$ with the $SU(2)_L$ stack $b$ and its orientifold image $b'$ are given by
\bea
(a \cap b, a \cap b') &=& (1,2) \quad {\rm if} \quad (-1)^{\tau ^a_0+\tau^b_0}=1 \label{ab1}\\
&=& (2,1) \quad {\rm if} \quad (-1)^{\tau ^a_0+\tau^b_0}=-1 \label{ab2}
\eea
thereby generating the required total of $3Q_L$ quark doublets (with $Y=\frac{1}{6}$). 
Similarly, using equation (247), the $U(1)$ stack $d$ and its orientifold image $d'$ have intersection numbers  
\beq
(a \cap d,a \cap d')=(0,0)
\eeq 
so that there are no quark-singlet states $q^c_L$ at these intersections. Choosing $y_c=-\frac{1}{2}$ in (\ref{Y}), the Higgs doublet $H_u$ with $Y=\frac{1}{2}$ arises at the intersection of $b$ with the  $U(1)$ stack $c$,  while $H_d$ with $Y=-\frac{1}{2}$ arises at the intersection with its  orientifold image $c'$:
\beq
(b \cap c,b \cap c')=(1,1)
\eeq 
The quark singlets arise at the intersections of $a$ with $c$ and $c'$
\beq
(c \cap a,c' \cap a)=(3,3)
\eeq
the former giving $3u^c_L$ and the latter $3d^c_L$. First, consider the case described by (\ref{ab2}). $u$-quark mass terms arising from the two $Q_L$ states at $a \cap b$ require the coupling of the states at $a \cap b$, $b \cap c$ and $c \cap a$, which is allowed by the conservation of $Q_a,Q_b$ and $Q_c$. However, the $u$-quark mass term arising from the  $Q_L$ state at $a \cap b'$ requires the coupling of the states at $a \cap b'$, $b \cap c$ and $c \cap a$, which is allowed by the conservation of $Q_a$ and $Q_c$, but {\em not} by $Q_b$, since the product has $\Delta Q_b=2$. 
Similarly, only two $d$-quark mass terms are allowed by conservation of $Q_b$. The alternative choice described by (\ref{ab1}) allows only one quark mass term for both $u$- and $d$-type quarks.

We also have
\bea
(d' \cap b,d' \cap b')=(1,2) && {\rm if} \quad (-1)^{\tau ^b_0+\tau ^d_0}\chi=-1=(-1)^{\tau ^a_0+\tau ^b_0} \label{bd1}\\
=(2,1) && {\rm if} \quad (-1)^{\tau ^b_0+\tau ^d_0}\chi=1=(-1)^{\tau ^a_0+\tau ^b_0} \label{bd2}
\eea
which generate the required total of $3L$ lepton doublets (with $Y=-\frac{1}{2}$), while the lepton singlets arise from 
\beq
(c' \cap d',c \cap d')=(3,3)
\eeq
the former giving  the $3\ell_L^c$ charged lepton singlets, and the latter the $3\nu_L^c$ the neutrino singlet states.
For the case (\ref{ab2}) under consideration, equation (\ref{bd1}) gives the location of the lepton doublets. 
 The charged lepton  mass term  arising from the lepton doublet at $d' \cap b$ require the coupling of the states at $d' \cap b$, $b \cap c'$ and $c' \cap d'$, which is allowed by the conservation of $Q_b,Q_c$ and $Q_d$. However, the  charged lepton  mass terms arising from the two lepton doublets at $d' \cap b'$ require  couplings that again have $\Delta Q_b=2$. Similarly, only one neutrino mass term is allowed by conservation of $Q_b$. The alternative choice described by (\ref{bd2}) allows two lepton mass term for both charged leptons and neutrinos.
Thus, at the perturbative level, after electroweak symmetry breaking,  we either have two massive quark generations and one massive lepton generation, or {\it vice versa}. In the model discussed in reference \cite{Bailin:2007va}, the same correlation is obtained.  

In both cases the missing couplings can only be provided by non-perturbative instanton effects. These  generate terms in the superpotential $W$ of the form
\beq
W \simeq \prod _i \Phi _i e^{-S_{\rm inst}}
\eeq
that { violate} the global $U(1)$ symmetries that survive after the Green-Schwarz mechanism breaks any  anomalous $U(1)$ gauge symmetry  \cite{Blumenhagen:2006xt};  here $\Phi _i$ are the (generally charged) matter superfields and $S_{\rm inst}$ is the action of the non-perturbative instanton. Such a term {\em is} allowed if and only if the gauge transformation of the matter field product $\prod_i\Phi _i $ under an anomalous $U(1)$ gauge transformation is cancelled by the transformation of the exponential factor induced by the {\em shift} of ${\rm Im} \ S_{\rm inst}$ under the $U(1)$ transformation \cite{Akerblom:2007nh}. 
Under a $U(1)_{\kappa}$ gauge transformation, associated with the stack $\kappa$,  parametrised by $\Lambda_{\kappa}$, in which the 1-form vector potential $A_1^{\kappa}$ is shifted by 
 \beq
\delta A_1^{\kappa}=d\Lambda_{\kappa}
\eeq
the imaginary part ${\rm Im} \ S_{\rm inst}$ of the instanton action  (\ref{SE2}) is shifted by
\beq
\delta\left( {\rm Im} \ S_{\rm E2}\right)={\Lambda_{\kappa} }Q_{\kappa}(E_2)
\eeq
 where $Q_{\kappa}(E_2)$ is the $U(1)_{\kappa}$ charge of the instanton, given by
\beq
Q_{\kappa}(E_2)= -\Xi \cap N_{\kappa}(\kappa- \kappa')
\eeq
To repair the missing Yukawa couplings we require that
\bea
Q_b(E_2)&=&-2  \label{QbE2}\\
Q_a(E_2)=0&=&Q_c(E_2)=Q_d(E_2) \label{Qacd}
\eea
The general form of $\Xi$ is given in equation (\ref{Xigen}) Then, using our solution for the $SU(2)_L$ stack on the {\bf AAA} lattice given in Table 1 and equation (66) of reference  \cite{Bailin:2008xx}, it follows from (\ref{QbE2}) above that
\beq
(-1)^{\tau ^b_0+1}[z_1+(-1)^{\tau ^b_2}z_5]=1 \label{Qbe}
\eeq
so that $z_1$ or $z_5$, but not both, are odd. 
We also require, as in (\ref{Qacd}), that the instanton has zero charge with respect to the other $U(1)$ charges.  For $Q_c$ this is guaranteed, since $c=c'$. Further, since $a-a'=d'-d$ in our solution, $Q_a(E2)=0$ ensures that $Q_d(E2)=0$. Thus, there is just one further constraint, which yields
\beq
 2Z_1-Z_3-Z_6+(-1)^{\tau^a_0}[z_1+(-1)^{\tau^a_2}z_6]=0 \label{Qa}
\eeq
It follows  from (\ref{Qbe}) that $\Xi$ is of $d_1$-type, as defined in equation (\ref{dtheta}), and it is easy to find solutions with all of the desired properties. For example
\beq
\Xi=\frac{1}{2}(\rho_1-\rho_3-\rho_4+\rho_6)+\frac{1}{2}(-1)^{\tau ^a_0+1}\left[\epsilon _1 +(-1)^{\tau ^a_2}\epsilon_6-\tilde{\epsilon} _1 -(-1)^{\tau ^a_2}\tilde{\epsilon}_6\right] \label{Xid1}
\eeq
with 
\beq
\tau ^a_0=\tau ^b_0 \bmod 2
\eeq
The above solution gives
\beq
(\Xi \cap b, \Xi \cap b')=(-1,-2)
\eeq
Thus the required total instanton charge (\ref{QbE2}) derives from one (massless) particle at the intersection of $\Xi$ with $b$, and { two}  at the intersections of $\Xi$ with $b'$.  To repair the missing $u$-quark Yukawa, for example, we need a 5-point coupling in which both $b$ and $b'$ intersect the fractional 3-cycle $\Xi$ of the instanton:
\beq
(a \cap b')(b' \cap \Xi)(\Xi \cap b)(b \cap c) (c \cap a) \label{xiabc}
\eeq
Since $\Xi \cap b=-1$,  we should interpret it as one intersection with $Q_b=+1$, rather than -1 intersections with $Q_b=-1$.  However, since $b' \cap \Xi=2$ is positive,   the coupling (\ref{xiabc}) does {\em not} then conserve $Q_b$, and we conclude that we can{\em not} repair the Yukawa with this E2-instanton.
Further, equation (\ref{QbE2}) requires that $\Xi \cap b-\Xi \cap b'=1$ which can only be satisfied with non-zero $\Xi \cap b$ and $\Xi \cap b'$ when they have the {\em same} sign, as in the above solution.
 Consequently $\Xi \cap b$ and $b' \cap \Xi$ cannot have the same sign in any of the solutions, and they therefore contribute zero to the total $Q_b$ charge in (\ref{xiabc}).  Hence we cannot repair the Yukawa with any of the single E2-instanton solutions of the constraints. The same conclusion follows for the model discussed in reference \cite{Bailin:2007va}, as well as for the models on the {\bf BAA} lattice given in Table 6 of reference \cite{Bailin:2008xx}.

\section{Conclusions}
All of the models that we have considered have the attractive feature that they have the spectrum of the supersymmetric \SM, including a single pair of Higgs doublets, plus
three right-chiral neutrino singlets.
In the presence of the previously derived non-zero background field strength $m_0H_3^{bg}$ they are also free of RR tadpoles, and therefore constitute consistent string-theory models.  We showed in \S \ref{susy} that this background field also stabilises {\em one} of the axion moduli. Further, we found  that it is easy to choose a non-zero background field strength $F_4^{bg}$ that stabilises the K\"ahler and complex-structure moduli associated with the supersymmetric mininima  at values within the K\"ahler cone in which the supergravity approximation is valid. In \S \ref{nonsusy} we showed that there are also {\em non}-supersymmetric stationary points of the effective potential, and in \S \ref{stab} we determined the parameter ranges in which these are stable minima. The stabilisation of {\em all} of the axion moduli can only be achieved by the use of non-perturbative instanton effects, and these were discussed in \S \ref{se2}. In principle, such  effects might also restore the missing quark and lepton Yukawa couplings to the Higgs doublets that are needed to generate masses when the electroweak symmetry is spontaneously broken.  However, we also showed that this does not happen for the particular models of interest here.


\vspace{1cm}
\noindent {\Large{\bf Acknowledgements}}
\vspace{0.5cm}

\noindent It is a pleasure to thank Joe Conlon and Andrei Micu for very helpful conversations, and also Gabriele Honecker and Timm Wrase for informative correspondence.
\vspace{1cm}

\appendix{\bf \LARGE Appendix A: \quad The Freed-Witten anomaly}

\vspace{1cm}
We need to assess whether any of our stacks $\kappa=a,b,c,d$ gives a non zero Freed-Witten anomaly
\beq
\Delta_{\kappa} \equiv H_3^{bg} \wedge [\kappa]
\eeq
where the background flux $H_3^{bg}$ in general has the form given in (\ref{H3bg}), and $[\kappa]$ is the 3-form that is the Poincar\'e dual of the fractional 3-cycle $\kappa$. The general form for the bulk part is
\beq
[\Pi_{\kappa}^{\rm bulk}]=\sum _p A^{\kappa}_p \eta_p
\eeq
where $\eta_p \ (p=1,3,4,6)$ are the 3-forms that are the Poincar\'e duals of the bulk 3-cycles $\rho_p$, given in eqns (329) ... (332) of reference \cite{Bailin:2008xx}.
As noted previously, we need only consider the $\mathcal{R}$-even part of $[\kappa]$. On the {\bf AAA} lattice, 
\bea
[\Pi_{\kappa}^{\rm bulk}]+[\Pi_{\kappa}^{\rm bulk}]'&=&A_3^{\kappa}(\eta_1+2\eta_3)+(2A_4^{\kappa}-A_6^{\kappa})\eta _4  \\
&=&\frac{6}{R_1R_3R_5}[3A_3^{\kappa}(\sigma _0-\sigma _1-\sigma _2+\sigma_3)-(2A_4^{\kappa}-A_6^{\kappa})(\sigma _0+\sigma _1+\sigma _2+\sigma_3)] \nonumber \\
&&
\eea
where the terms on the right-hand side are defined in eqns (\ref{sig0}) ... (\ref{sig3}).
Then 
\beq
H_3^{bg} \wedge [\Pi_{\kappa}^{\rm bulk}]=-\frac{6i}{R_1R_3R_5\sqrt{{\rm Vol}_6}}[3A_3^{\kappa}(p_0+p_1)+(2A_4^{\kappa}-A_6^{\kappa})(p_1-p_0)]w_1 \wedge w_2 \wedge w_3
\eeq
where the (1,1)-forms $w_k \ (k=1,2,3)$ are defined in eqn (\ref{wk}). 
Similarly, the general form for the exceptional part is
\beq
[\Pi_{\kappa}^{\rm ex}]=\sum _j (\alpha^{\kappa}_j \chi_j+\tilde{\alpha}^{\kappa}_j \tilde{\chi}_j)
\eeq
where $\chi_j, \tilde{\chi}_j \ (j=1,4,5,6)$ are the Poincar\'e duals of the exceptional 3-cycles $\epsilon_j, \tilde{\epsilon}_j$; these are defined in eqns (369) and (370) of reference \cite{Bailin:2008xx}. 
Then, on the {\bf AAA} lattice, 
\bea
[\Pi_{\kappa}^{\rm ex}]+[\Pi_{\kappa}^{\rm ex}]'&=&\sum _j (2\tilde{\alpha}^{\kappa}_j -\alpha^{\kappa}_j)\tilde{\chi}_j\\
&=&\sum _j \frac{R_3}{2\pi \hat{\lambda}_j^{2}\alpha {\rm Vol}_2}(2\tilde{\alpha}^{\kappa}_j -\alpha^{\kappa}_j)(\omega _j-\alpha \tilde{\omega}_j)
\eea
where $\omega _j$ and $\tilde{\omega}_j$ are defined in eqns (\ref{omj}) and (\ref{tilomj}).
It follows that
\beq
H_3^{bg} \wedge [\Pi_{\kappa}^{\rm ex}]=\sum _{i,j}\frac{iR_3}{2\pi^2\hat{\lambda}_j^4{\rm Vol}_2^{3/2}}(2\tilde{\alpha}^{\kappa}_j -\alpha^{\kappa}_j)P_je_{(i,j)} \wedge e_{(i,j)} \wedge w_2
\eeq
Here $e_{(i,j)}$ are the localised (1,1)-forms defined in (\ref{eij}).

Consider first the solution on the {\bf AAA} lattice  given in \S 5.1 of reference \cite{Bailin:2008xx}, derived from the fourth solution in Table 1 of that paper. For the bulk parts of $\kappa= a,b,c,d$ respectively we have
\beq
3A_3^{\kappa}(p_0+p_1)+(2A_4^{\kappa}-A_6^{\kappa})(p_1-p_0)=(6p_1,0,0,-6p_1)  
\eeq
so that cancellation of (the bulk part of) the Freed-Witten anomaly $\Delta^{\rm bulk}_{\kappa}$ for the $\kappa=a$ and $d$ stacks requires that $p_1=0$. It follows from (\ref{p0p1a}) that this in turn requires that $n_3=3n_6$ and it is evident from the discussion following eqn (\ref{p0p1a}) that this is {\em not} satisfied by any of our solutions. 
Thus $\Delta^{\rm bulk}_{\kappa}\neq 0$ for the stacks $\kappa=a$ and $d$. Correspondingly, for the exceptional parts we find 
\bea
2\tilde{\alpha}^{\kappa}_j -\alpha^{\kappa}_j&=&-3t^a_0(1,0,0,t^a_2)  \\
&=&2t^b_0(1,0,t^b_2,0) \\
&=&(0,0,0,0) \\
&=&3t^a_0(1,0,0,t^a_2) 
\eea
The localisation of the (1,1)-forms $e_{(i,j)}$ means that the cancellation of (the exceptional part of) the Freed-Witten anomaly $\Delta^{\rm ex}_{\kappa}$ for  $\kappa=a,b$ and $d$ requires that $P_1=0=P_5=P_6$, and hence that $\hat{n}_1=0=\hat{n}_5=\hat{n}_6$. We have already noted that tadpole cancellation requires that $|n_0\hat{n}_j|=12$ for $j=4,6$ so the last of these can{\em not} be satisfied; with the choice $t^c_1=1$, though, we can satisfy the first of these. Thus, $\Delta^{\rm ex}_{\kappa}\neq 0$ at least for the stacks $\kappa=a$ and $d$.  A very similar analysis, with the same conclusions, applies to the other solutions derived from Table 1. 
The solutions on the {\bf BAA} lattice are discussed in \S 6.2, and derive from Table 6 of reference \cite{Bailin:2008xx}. We again find in all cases that  $\Delta^{\rm bulk}_{\kappa}\neq 0$ for the stacks $\kappa=a$ and $d$.


\newpage
\appendix{\bf \LARGE Appendix B: \quad Positive definiteness of the axionic fluctuations}

\vspace{1cm}

It is clear from its definition  that $D(a)\equiv (2 \gamma ^2+1) a^2-3$ is negative for $a_-<a <a_+$, where
\beq
a_{\pm} \equiv \pm \sqrt{\frac{3}{2\gamma ^2+1}} 
\eeq
and positive elsewhere. For future information, it is easy to verify that 
\bea
a_1\equiv \frac{1}{4\gamma^2}\left(1-\sqrt{1+8\gamma^2}\right)&>&a_- \quad  \forall \gamma \\
\frac{1}{4\gamma^2}\left(1+\sqrt{1+8\gamma^2}\right)&>&a_+\quad {\rm for}  \quad \gamma^2 <1 \\
\frac{1}{4\gamma^2}\left(1+\sqrt{1+8\gamma^2}\right)&<&a_+ \quad {\rm for} \quad \gamma^2 >1 
\eea
The analysis of $N(a)=(4\gamma ^2-1) a^4+4\gamma ^2 a^3+1 $, defined in equation (\ref{Ndef}), is more complicated. For all values of $\gamma^2$ it has a root at $a=-1$, a saddle point at $a=0$, and one other stationary point at 
\beq
a=a_D \equiv \frac{3\gamma ^2}{1-4\gamma ^2}
\eeq
When $0 <\gamma ^2 <1/4$ this stationary point is at a positive value of $a$ and is a maximum. In this case, $N(a)$ is positive for $a_4 <a<  a_5$, and negative elsewhere; here $-1=a_4<a_1<0$ and $ a_5>a_D>0$ is the  (positive) root $\alpha(\gamma^2)$ of the cubic factor in equation (\ref{N2}); in the special case $\gamma=0$, for example, $a_5=1$. 
It is easy to verify that $N(a_-) \leq 0$ (actually for all values of $\gamma^2$, with equality only when $\gamma ^2=1$); thus $a_-<a_4<a_1$. Further, $N(a_+)$ is negative for $0< \gamma ^2 \lesssim 0.1255$, vanishes when $\gamma ^2 \simeq 0.1255$, and is positive for all other values of $\gamma ^2$; it follows that $a_5<a_+$ for $0< \gamma ^2 \lesssim 0.1255$, but $a_+<a_5$ for $ 0.1255\lesssim \gamma ^2 <1/4 $. 
Alternatively, when $\gamma ^2>1/4$, $a_D$ is negative  and $N(a_D)$ is a minimum. In this case, $N(a)$ is negative for $a_4 <a<  a_5$, and positive elsewhere, and now both $a_4$ and $a_5$ are negative roots of $N(a)$, with $a_4 <a_D <a_5$; for $\gamma^2 <1$ the position of the stationary point satisfies $a_D<-1$, whereas for $\gamma^2 >1$ we find $a_D>-1$. Thus when $\gamma^2 <1$, $-1=a_5<a_1$ and $a_4$ is the root of the cubic in equation (\ref{N2}), whereas for $\gamma^2 >1$,  $-1=a_4<a_1$ and $a_5$ is the root of the cubic. Obviously, $a_+>a_5$ for values of $\gamma ^2$ in this range.
These considerations lead us to consider  three ranges of values for $\gamma ^2$, with the  signs of the functions given in the associated Tables.
\begin{itemize}
\item $0<\gamma ^2 \lesssim 0.1255 $ 
\begin{table}[h!]
 \begin{center}
\begin{tabular}{||c||c|c|c|c||} \hline \hline
Region & $a$ & $N(a)$ & $D(a)$ & $d_6(a)$  \\ \hline \hline 
I & $a < a_-$ &- & + & -  \\ \hline
II & $a_-<a<-1$ & -& - & \\ \hline
III & $-1<a<a_5$ & + & -& + \\ \hline
IV & $a_5<a <a_+$& -&-&  \\ \hline
V &$a_+<a$ & -& +& -   \\ \hline \hline
\end{tabular}
\end{center} 
\caption{ \label{T1} Signs of the functions $N(a), D(a), d_6(a)$ when $0<\gamma ^2 <0.1255 $}
 \end{table} 

We need to identify the regions in which {\em both} $d(a)$ and $d_6(a)$ are positive. With $a_1$ defined in equation (\ref{x2}), we note that 
$D(a_1)$ is negative for any value of $\gamma ^2 $. Similarly,   $N(a_1)$  is positive (actually for any value of $\gamma ^2$). It follows that  $a_1$ is in region III of Table \ref{T1}; this is consistent with the observation above that $a_4<a_1$ in this case. 
 From equation (\ref{gamsmall}) we see that $a_2=1$ for values of $\gamma ^2$ in this range. Since $D(1)=2(\gamma^2-1)<0$, in this case,  and $N(1)=8\gamma^2>0$, it follows that $a_2$ is also in region III of Table \ref{T1}. Finally, using the value of $a_3$ given in (\ref{gamsmall}), we find that $D(a_3)$ is positive, (actually for any  $\gamma ^2 <1$;  it vanishes when  $\gamma^2 =1$, and is negative for all other values.)
For values of $\gamma^2 \lesssim 0.1915$,  $N(a_3)$ is negative; (it vanishes when  $\gamma^2 \simeq 0.1955$, and is positive for all other values.) It follows that $a_3$ is in region V of Table \ref{T1} in which $d_6(a)$ is negative. Thus the only region in which both $d(a)$ and $d_6(a)$ are positive  is $a_1<a<a_2=1$,  and this is the case for all values of $\beta ^2$; note that this range does not require the solution of the cubic.
\item $0.1255<\gamma ^2 <\frac{1}{4} $ \quad 

The properties of the functions given above show that in this case $a_1$ is in region III of Table \ref{T2}, as is $a_2$, and if $\gamma ^2 \lesssim 0.1955$ then $a_3$ is in region V; otherwise it is region IV. Thus, if $\gamma ^2 \lesssim 0.1955$, the only region in which both $d(a)$ and $d_6(a)$ are positive   is again $a_1<a<a_2=1$, and as before this is the case for all values of $\beta ^2$.
However, in the case 
$\gamma ^2 \gtrsim 0.1955$,  $N(a)$ and $D(a)$ are both positive in
 the region $a_3 <a <a_5$, so that both $d(a)$ and $d_6(a)$ are positive here too,  provided that $\beta ^2<N(a)/D(a)$. The determination of $a_5$ requires the solution of the cubic, which we discuss below. 
\begin{table}[h!]
 \begin{center}
\begin{tabular}{||c||c|c|c|c||} \hline \hline
Region & $a$ & $N(a)$ & $D(a)$ & $d_6(a)$  \\ \hline \hline 
I & $a < a_-$ &- & + & -  \\ \hline
II & $a_-<a<-1$ & -& - & \\ \hline
III & $-1<a<a_+$ & + & -& + \\ \hline
IV & $a_+<a <a_5$& +&+&  \\ \hline
V &$a_5<a$ & -& +& -   \\ \hline \hline
\end{tabular}
\end{center} 
\caption{ \label{T2} Signs of the functions $N(a), D(a), d_6(a)$ when $0.1255<\gamma ^2 <1/4 $}
 \end{table} 
\item $\gamma ^2 >\frac{1}{4} $ 

In this case we conclude that $a_1$ is in region IV of Table \ref{T3}, as is  $a_2$, and    $a_3$ is in region V.
Thus again  both $d(a)$ and $d_6(a)$ are positive  in the region  $a_1<a<a_2$, and this is the case  for all values of $\beta ^2$.  
  As noted previously, there is a further region  $a >a_3$ in which positive-definitenesss is assured provided that $\beta ^2<N(a)/D(a)$. Since $N(a)$ grows with $a$ more rapidly than $D(a)$, the upper bound on $\beta^2$ grows with $a$. 
\begin{table}[h!]
 \begin{center}
\begin{tabular}{||c||c|c|c|c||} \hline \hline
Region & $a$ & $N(a)$ & $D(a)$ & $d_6(a)$  \\ \hline \hline 
I & $a < a_4$ &+ & + &   \\ \hline
II & $a_4<a<a_-$ & -& + & -\\ \hline
III & $a_-<a<a_5$ & - & -&  \\ \hline
IV & $a_5<a <a_+$& +&-&  +\\ \hline
V &$a_+<a$ & +& +&    \\ \hline \hline
\end{tabular}
\end{center} 
\caption{ \label{T3} Signs of the functions $N(a), D(a), d_6(a)$ when $\gamma ^2>1/4  $. For $\gamma^2<1$ $a_5=-1$, whereas for $\gamma^2>1$ $a_4=-1$.}
 \end{table} 
\end{itemize}


To solve the cubic we write $N(a)$, defined in equation (\ref{N2}) in the form
\beq
N(a)=(a+1)(4\gamma^2-1)C(a)
\eeq
 where $C(a)$ has the form
\beq
C(a)=a^3+c_2a^2+c_1a+c_0 \label{cubic}
\eeq
with
\beq
c_2=\frac{1}{4\gamma^2-1}=-c_1=c_0 \label{adef}
\eeq
The roots of $C(a)$ are found by first changing variables from $a$ to $y$, where
\beq
a=y-\frac{1}{3}c_2 \label{ay}
\eeq
to cast it in the canonical form
\beq
y^3+py=q \label{cancub}
\eeq
with
\bea
p &\equiv& c_1-\frac{1}{3}c_2^2 =\frac{2(1-6\gamma^2)}{3(4\gamma^2-1)^2} \label{pdef}\\
q &\equiv& \frac{1}{27}(9c_1c_2-27 c_0-2 c_2^3)=\frac{4(45\gamma^2-108\gamma^4-5)}{27(4\gamma^2-1)^3} \label{qdef}
\eea
Equation (\ref{cancub}) is solved by making Vieta's substitution
\beq
y=w-\frac{p}{3w} \label{yw}
\eeq
so that
\bea
w^3&=&\frac{q}{2} \pm \sqrt{\frac{q^2}{4}+\frac{4p^3}{27}} \\
&=&\frac{2(45\gamma^2-108\gamma^4-5)}{27(4\gamma^2-1)^3} \pm \frac{2}{(4\gamma^2-1)^2}\sqrt{\frac{27\gamma^4-10\gamma^2+1}{27}} \label{w3}
\eea
Thus we can solve for  any value of $\gamma^2\neq 1/4$. Consider first the plus sign in equation (\ref{w3}). For $0 \leq \gamma^2<1/4$, the root $\alpha(\gamma^2)$ of the cubic $C(a)=0$ increases monotonically from $\alpha(0)=1$ with $\alpha(\gamma^2)\rightarrow +\infty$ as $\gamma ^2 \rightarrow 1/4$ from below. Thus in this case $a_5>1$. For $\gamma^2 >1/4$, the root $\alpha(\gamma^2)$ is negative and monotonically increasing as $\gamma^2$ increases, with $\alpha(\gamma^2)\rightarrow - \infty$ as $\gamma ^2 \rightarrow 1/4$ from above, and $\alpha(\gamma^2)\rightarrow 0$ as $\gamma ^2 \rightarrow \infty$. For the solution corresponding to the minus sign in (\ref{w3}) one has to be more careful because $w(1/6)=0=p(1/6)$ which makes the evaluation of $y$ undefined at that value of $\gamma^2$; one has to do the ranges $0 \leq \gamma^2 <1/6$ and  $1/6 < \gamma^2 <1/4$   separately, even though there's nothing special about the cubic for this value of $\gamma^2$. 
However, the conclusion is that both signs give the same value of the root $\alpha(\gamma^2)$ for any given value of $\gamma^2$. 

The following two examples illustrate the procedure.
\begin{itemize}

\item $\gamma^2=0.2$

Since $\gamma^2>0.1955$,  we expect positive-definiteness for some values of $a$ in region IV of Table \ref{T2}, besides those in $-0.7656=a_1<a<a_2=1$ in region III. For this value of $\gamma^2$ equations (\ref{gamsmall}), (\ref{w3}), (\ref{yw}) and (\ref{ay}) yield
\beq
a_3=3.266 \quad {\rm and} \quad a_5=4.074
\eeq
so for values $a$ in the range $a_3<a<a_5$,  $d(a)$ is positive, and so are  $N(a)$ and $D(a)$. Then it follows from (\ref{Ddef}) that $d_6(a)$ too is positive  for values of $\beta^2$ satisfying
\beq
\beta^2 <\frac{N(a)}{D(a)}=\frac{-a^4+4a^3+5}{7a^2-15}
\eeq
In this range the upper bound on $\beta^2$ is approximately linear, starting at $\beta^2 \lesssim 0.51$ when $a=a_3$ and decreasing to zero at $a=a_5$, where $N(a)$ vanishes.
\item $\gamma ^2=2$

The solution of the cubic is not needed in this case. Besides the range $-0.3904=a_1<a<a_2=0.6404$ in  region IV of Table \ref{T3}, both $d_6(a)$ and $d(a)$ are positive in the range $a>a_3=1$ in region V, provided that $\beta^2<\frac{7a^4+8a^3+1}{5a^2-3}$. Positive-definiteness is assured for all values of $a$ in this range when $\beta ^2 \lesssim 6.985$, whereas larger values of $\beta^2$ are only allowed for larger values of $a$.

\end{itemize}



\end{document}